%% file: mnras_template.tex
\documentclass[fleqn,usenatbib]{mnras}

\usepackage{newtxtext,newtxmath}

\usepackage[T1]{fontenc}
\usepackage{ae,aecompl}

\usepackage{graphicx}	
\usepackage{amsmath}	
\usepackage{amssymb}	
\usepackage{float}

\newcommand{\lm}{$\mathrm{L_{max}}$}
\newcommand{\Nif}{$\rm ^{56}Ni$\,} 
\newcommand{\Cof}{$\rm ^{56}Co$\,} 
\newcommand{\Fef}{$\rm ^{56}Fe$\,} 
\newcommand{\Mni}{$\rm M_{^{56}Ni}$}
\newcommand{\g}{$\gamma$}
\newcommand{\ejm}{$M_{\rm ej}$\,}
\newcommand{\mch}{$M_{\rm ch}$\,}

\usepackage{color}

\title[iPTF16abc and the population of Type Ia supernovae]{iPTF16abc and the population of Type Ia supernovae: Comparing the photospheric, transitional and nebular phases}

\author[S. Dhawan et al.]{S. Dhawan,$^{1}$\thanks{E-mail: suhail.dhawan@fysik.su.se}
M. Bulla,$^{1}$
A. Goobar,$^{1}$
R. Lunnan,$^{2}$
J. Johansson,$^{3}$
C. Fransson,$^{2}$
\newauthor
S.R. Kulkarni$^{4}$, S. Papadogiannakis$^{1}$, A.A.Miller$^{5,6}$
\\
$^{1}$ The Oskar Klein Centre, Physics Department, Stockholm
  University, SE 106 91 Stockholm, Sweden\\
  $^{2}$ Department of Astronomy, The Oskar Klein Centre, Stockholm University, Alba Nova University Centre, SE-106 91 Stockholm, Sweden\\
$^{3}$ Department of Physics and Astronomy, Division of Astronomy and Space Physics, Uppsala University, Box 516, SE 751 20 Uppsala, Sweden \\
$^{4}$ Division of Physics, Mathematics and Astronomy, California Institute of Technology, Pasadena, CA 91125, USA\\
$^{5}$ Center for Interdisciplinary Exploration and Research in Astrophysics
(CIERA) and Department of Physics and Astronomy,
Northwestern University, 2145 Sheridan Road, Evanston, IL 60208,
USA\\
$^{6}$ The Adler Planetarium, Chicago, IL 60605, USA\\
}

\date{Accepted XXX. Received YYY; in original form ZZZ}

\pubyear{2017}

\hypersetup{draft}

\begin{document}
\label{firstpage}
\pagerange{\pageref{firstpage}--\pageref{lastpage}}
\maketitle

\begin{abstract}
Key information about the progenitor system and the explosion mechanism of Type Ia supernovae (SNe~Ia) can be obtained from early observations, within a few days from explosion. iPTF16abc was discovered as a young SN~Ia with excellent early time data. Here, we present photometry and spectroscopy of the SN in the nebular phase. A comparison of the early time data with a sample of SNe~Ia shows distinct features, differing from normal SNe~Ia  at early phases but similar to normal SNe~Ia at a few weeks after maximum light (i.e. the transitional phase) and well into the nebular phase.    
The transparency timescales ($t_0$) for this sample of SNe~Ia range between $\sim$ 25 and 41 days indicating a diversity in the ejecta masses. $t_0$ also weakly correlates with the peak bolometric luminosity, consistent with the interpretation that SNe with higher ejecta masses would produce more \Nif. 
Comparing the $t_0$ and the maximum luminosity, \lm\, distribution of a sample of SNe~Ia to predictions from a wide range of explosion models we find an indication that the sub-Chandrasekhar mass models span the range of observed values. However, the bright end of the distribution can be better explained by Chandrasekhar mass delayed detonation models, hinting at multiple progenitor channels to explain the observed bolometric properties of SNe~Ia. iPTF16abc appears to be consistent with the predictions from the \mch models. 
\end{abstract}

\begin{keywords}
supernova: general-supernova: individual (iPTF16abc)
\end{keywords}



\section{Introduction}
Type Ia supernovae (SNe~Ia) have long been linked to the explosion of a C/O white dwarf (WD) in a binary system \citep{Hoyle1960}, which has been confirmed by observed limits on the progenitor \citep{Nugent2011,Bloom2012}. However, there is still heated debate about the fundamental physical properties of the system, e.g. the mass of the progenitor, the nature of the companion and the mechanism of the explosion \citep[see,][for a review]{Hillebrandt2013,Maoz2014}. 
Many attempts to answer these open questions regarding the physics of SNe~Ia have concentrated on observations near maximum light. These observations have been critical to derive global parameters for SNe~Ia, e.g. synthesized \Nif\, mass, total ejecta mass \citep{Stritzinger2006a,Scalzo2014,Dhawan2016,Dhawan2017a}. 
However, there is a wealth of information available in observations shortly after explosion as well as at late times \citep[$\sim$ a year after maximum, i.e. the nebular phase, e.g.][]{Maguire2016,Graham2017}. Early time observations can shed light on the interaction between the SN ejecta and its companion \citep{Kasen2010}, and can be used to constrain the size of the companion. The signature  of such an interaction can be seen as a sharp excess in the UV and blue flux at early epochs \citep[e.g.][]{Cao2015}, however, these pulses can be interpreted differently as in \citet{Kromer2016} and \citet{Noebauer2017}. Very early time observations of SNe~Ia \citep{Zheng2013,Zheng2014,Goobar2014,Goobar2015,Marion2016,Hoss2017,Miller2017,Jiang2017} have shown a diversity in their behaviour shortly after explosion and are a rich source of information regarding the progenitor system and the explosion mechanism.

 iPTF16abc was discovered shortly after explosion and showed some interesting early time features. It has a linear rise for the first three days after the time of first light, blue colours at early times compared to normal SNe \citep[e.g. SN~2011fe][]{Nugent2011}, strong carbon features in early spectra that disappear after $\sim$ 7 days  \citep{Miller2017} and the near absence of the Si II\, 6355\,\AA\, in the earliest spectrum. 
Here, we present a nebular spectrum of iPTF16abc and analyse the photospheric (i.e. pre-maximum), transitional (i.e. $\sim$ +30 to +100 days) and nebular ($\sim$ +300 days) phase properties of iPTF16abc in context of SNe in the literature. 

The observations of the early, photospheric phase mostly probe the outer layers of the ejecta, the late phase, when the $\gamma$-ray escape fraction increases \citep{Jeffrey1999,Stritzinger2006a}, is sensitive to the inner layers of the SN ejecta, which are dominated by iron group elements (IGEs). We, therefore, aim to answer if the features seen in the early phase also manifest at late epochs 
and hence, whether they are a result of only the composition of the outer layers of the ejecta or also due to the inner core. 

The structure of the paper is as follows. We present the dataset in Section~\ref{sec-data}. We compare the properties of iPTF16abc to a sample of SNe in Section~\ref{sec-res}. We discuss our results in Section~\ref{sec-disc} and conclude in Section~\ref{sec-conc}. 


\section{Data}
\label{sec-data}
The intermediate Palomar Transient Factory (iPTF)  reported
the discovery of iPTF16abc (IAU name: SN2016bln), located
170$''$ from the galaxy NGC 5221 at a redshift of $z=0.023$ and R.A., Dec = 13:34:45.492
+13:51:14.30 (J2000). The line-of-sight that has low Milky Way
reddening with $E(B-V)_{MW} = 0.028$ mag \citep{Schlafly2011}. The supernova was discovered on 4.4 April 2016 and the first detection was on 3.4 April 2016, the last non-detection was on 2.4 April 2016 \citep[see,][for details]{Ferretti2017}. The distance to the host galaxy was calculated using a value of $H_0$ = 70 kms$^{-1}$Mpc$^{-1}$ and standard cosmology (i.e. flat, $\Omega_m =0.3$) which corresponds to a distance of 100.24 $\pm$ 4.3 Mpc (using a peculiar velocity error on the redshift of 300 kms$^{-1}$), or a distance modulus of 35.00 $\pm$ 0.10 mag. Note that since the SN is in the linear part of the Hubble diagram ($z = 0.023$), the effect of the assumed cosmology, apart from $H_0$ is very small.

As part of a late phase follow-up of iPTF16abc, we obtained a spectrum with the Low Resolution Imaging Spectrometer (LRIS) on the Keck telescope at +342.4\,days. The spectrum was obtained as part of program C299 (PI: Kulkarni). It was reduced using the standard LRIS reduction pipeline, \texttt{lpipe}\footnote{http://www.astro.caltech.edu/~dperley/programs/lpipe.html}, written in \texttt{IDL}. A summary of the spectra is provided in Table~\ref{tab:spec_log}. Moreover, we also present photometry at similar epochs with DECam as part of the Dark Energy Camera Legacy Survey (DECaLS), reduced using standard PSF fitting photometry routines in \texttt{python}. 

In this study, we also analyse the data presented in \citet{Miller2017} and \citet{Ferretti2017}. To this data, we add multi-band data for a sample of SNe~Ia from the literature. Since we analyse the multi-band and bolometric properties of a number of SNe~Ia, we require coverage from $u \rightarrow H$. We use data from the Carnegie Supernova Project \citep[CSP;][]{Contreras2010,Stritzinger2011} and the CfA supernova program \citep{Friedman2015}. 

\input{spectroscopic_log}

\input{dm15_vals}
\input{distances}

\section{Results}
\label{sec-res}

\subsection{Spectroscopy}

\begin{figure*}
\includegraphics[width=1\textwidth]{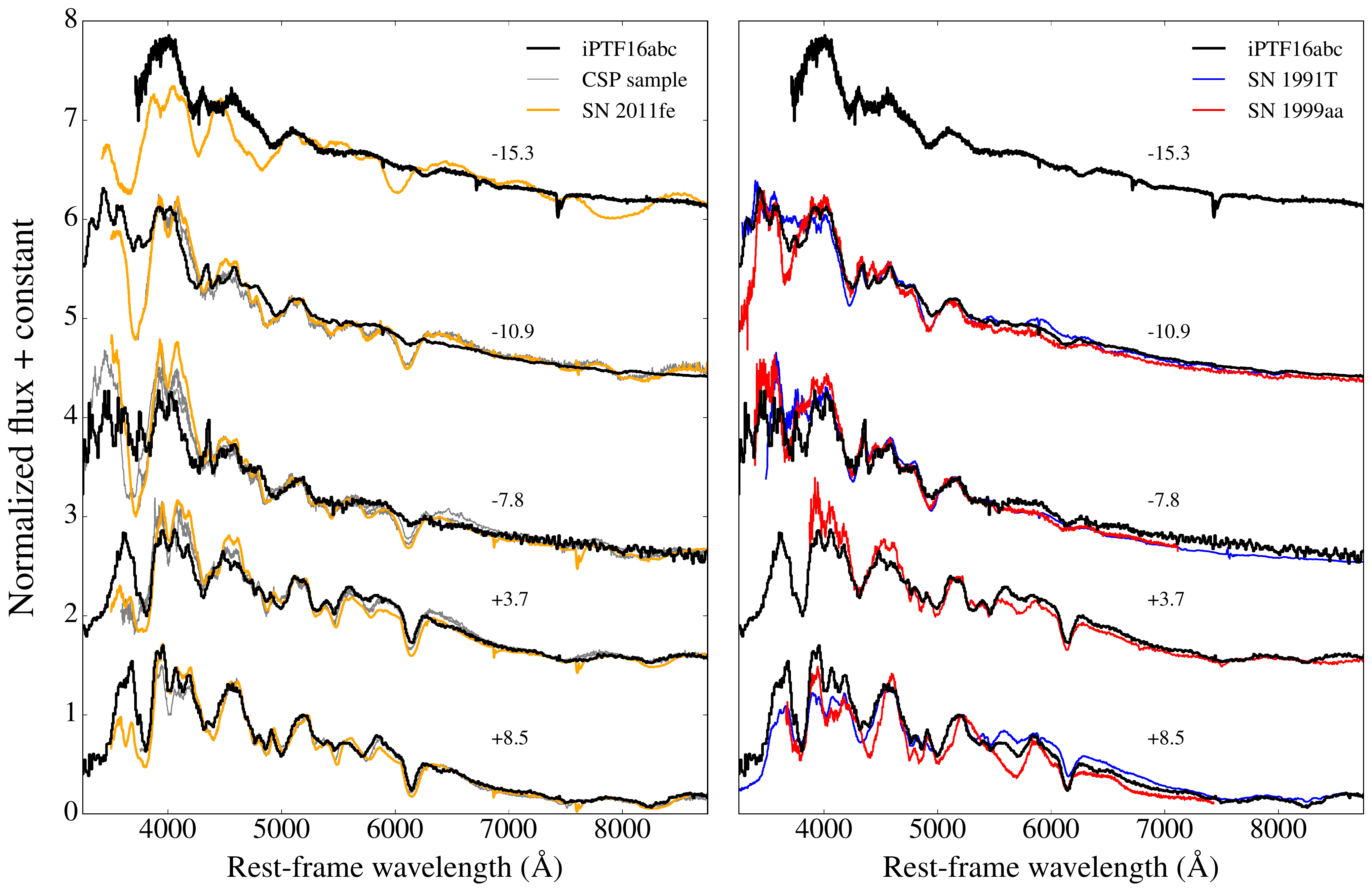}
\caption{{\emph Left:} The early time ($<$ +10 days) spectral time sequence of iPTF16abc (black) is compared to the spectra of SN 2011fe (orange), some normal SNe~Ia in the CSP sample of Table~\ref{tab:ebv_rv} (grey) {\emph Right:} The spectral time sequence is compared to SN~1991T and SN~1999aa at similar epochs ($\pm$2.5~d). Spectra in the left panel are normalized by their values at $\sim$~5200~\AA, the top four in the right panel by their values at $\sim$~5500~\AA{} and the last spectra by their values at $\sim$~5200~\AA.}
\label{fig:16abc_spectra}
\end{figure*}

\begin{figure*}
\includegraphics[width=1\textwidth]{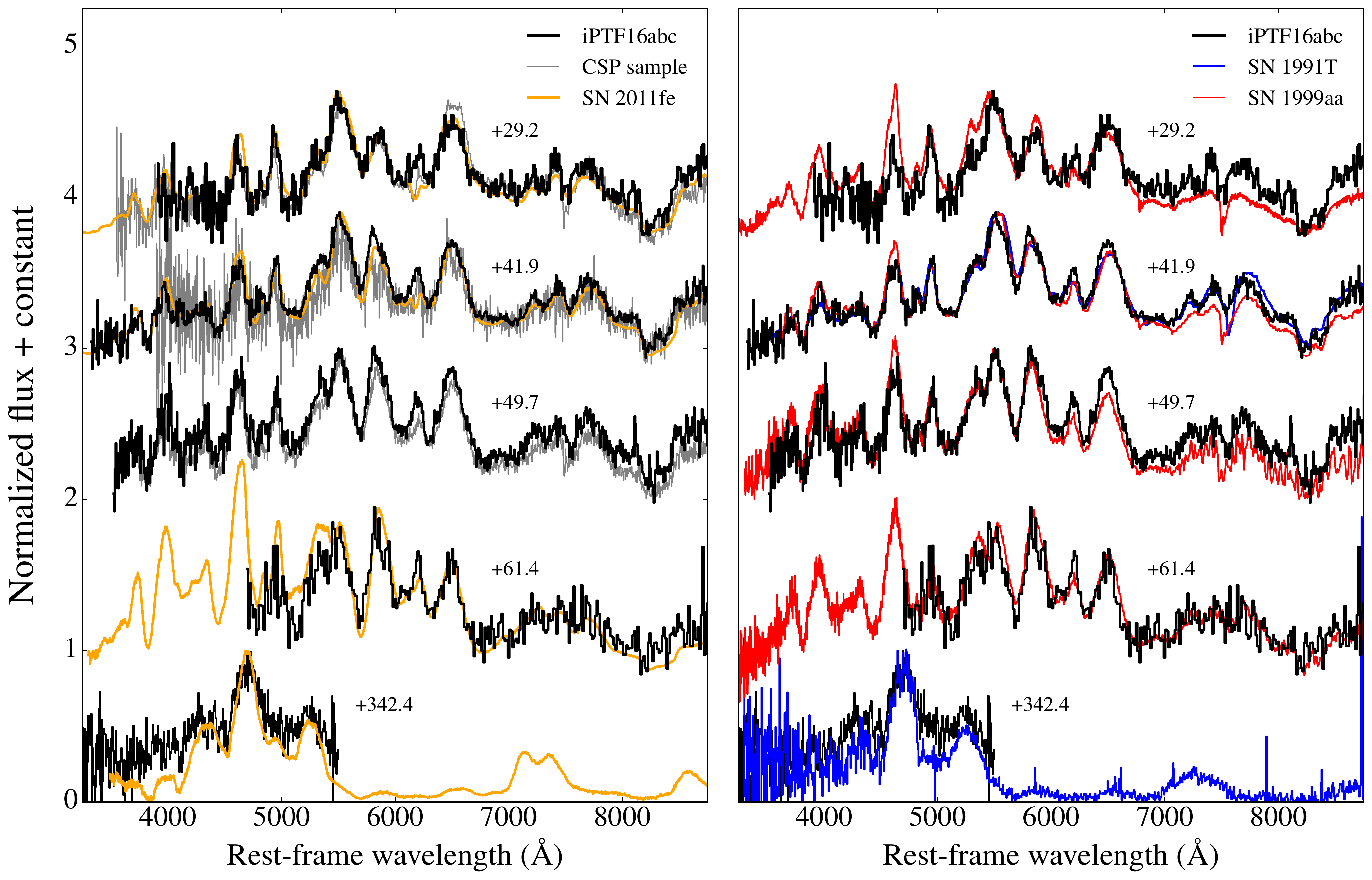}
\caption{ Same as Figure~\ref{fig:16abc_spectra} but for the transitional and the nebular phase. The top four spectra in the right panel are normalised to their values at $\sim$~5500~\AA{} and the last spectrum by its value at $\sim$~4700~\AA.}
\label{fig:16abc_spectra_late}
\end{figure*}

Figure~\ref{fig:16abc_spectra} shows observed spectra of iPTF16abc from $-$15.3 days to +8.5~d relative to $B-$band maximum and figure~\ref{fig:16abc_spectra_late} from +29.2\,d to +342.4\,d, compared to those of some normal SNe~Ia at similar epochs ($\pm$2.5~d). Specifically, the comparison sample comprises spectra for the well-studied and nearby SN~2011fe \citep{Nugent2011,Pereira2013,Maguire2014,Mazzali2014,Taubenberger2015} together with those for normal SNe~Ia in the CSP sample of Table~\ref{tab:ebv_rv} that are characterised by $E(B-V)<$0.2~mag and no high-velocity features (i.e. SN2004ey, SN2005el, SN2005na, SN2006ax and SN2008hv, \citealt{Folatelli2013}).  We also plot the available spectra for the overluminous SN~1991T \citep{Filippenko1992,Phillips1992,Ruizlapuente1992,Gomez1998,Silverman2013} and SN~1999aa \citep{Garavini2004,Matheson2008,Silverman2013} in figures~\ref{fig:16abc_spectra} and \ref{fig:16abc_spectra_late}.

Spectra of iPTF16abc before maximum show distinct peculiarities compared to the bulk of normal SNe~Ia at these early epochs \citep{Miller2017}. The most striking difference is found across the Si\,{\sc ii}\,$\lambda6355$ and Ca\,{\sc ii} near-IR triplet features around 6000 and 8000~\AA{}, respectively. Both these features are usually strong in pre-maximum spectra of normal SNe~Ia, while they are much weaker in iPTF16abc.   A \texttt{SNID} classification of the -11\,d spectrum in Figure~\ref{fig:16abc_spectra} points to a best match with SN~1999aa, an overluminous, peculiar SN~Ia \citep{Garavini2004}, which is consistent with the deep Ca H\&K feature in iPTF16abc that is not seen in SN~1991T (Figure~\ref{fig:16abc_spectra}).  The early time ($<$ -10 days) colours for iPTF16abc are significantly bluer \citep[see][]{Miller2017} and the colour evolution is significantly flatter than other overluminous SNe with such early observations \citep[e.g. SN2012fr;][]{Contreras2018}. We emphasize, however, that our results are independent of the classification of iPTF16abc as a 91T-like, 99aa-like or normal SN~Ia.
As we will discuss in Section~\ref{sec-disc}, this behaviour is consistent with the suggestion from \citet{Miller2017} that strong mixing could have occurred in the ejecta of iPTF16abc. 

Despite the pronounced differences seen at early-epochs, the spectral time evolution of iPTF16abc from peak brightness to about a year after is remarkably similar to the one observed in normal SNe~Ia  which have been shown to be similar to 91T-like/99aa-like SNe \citep{Filippenko1992,Garavini2004}, also consistent with the \texttt{SNID} classification from the post maximum spectrum, presented in \citet{Miller2017}. Good agreements with the comparison spectra are found at all epochs both in terms of colours \citep[see also][]{Miller2017} and velocities/ strength of individual features. The similarities between iPTF16abc and the comparison sample extend up to  transitional and nebular phases ($\gtrsim$~30~d), when the inner regions of the SN ejecta are probed.

\subsubsection{Nebular Phase}
Connections between SN~Ia properties at early times and those at epochs $\gtrsim$~150~d have been proposed, including correlations between the strength of the Fe $\lambda$4700 feature and $\Delta m_{15,B}$ \citep[][but see also \citealt{Blondin2012}]{Mazzali1998} and between nebular line velocities and colours at peak \citep{Maeda2011}. Measuring late-time properties of iPTF16abc can improve our understanding of the observed peculiarities at early times. We obtained a spectrum of iPTF16abc at +342.4\,days close to a year after maximum light. In Figure~\ref{fig:16abc_spectra_late} we qualitatively compare the spectrum to SN~2011fe, SN~1991T and SN~1999aa in the nebular phase and find no striking differences. 

\begin{figure}
\includegraphics[width=.5\textwidth]{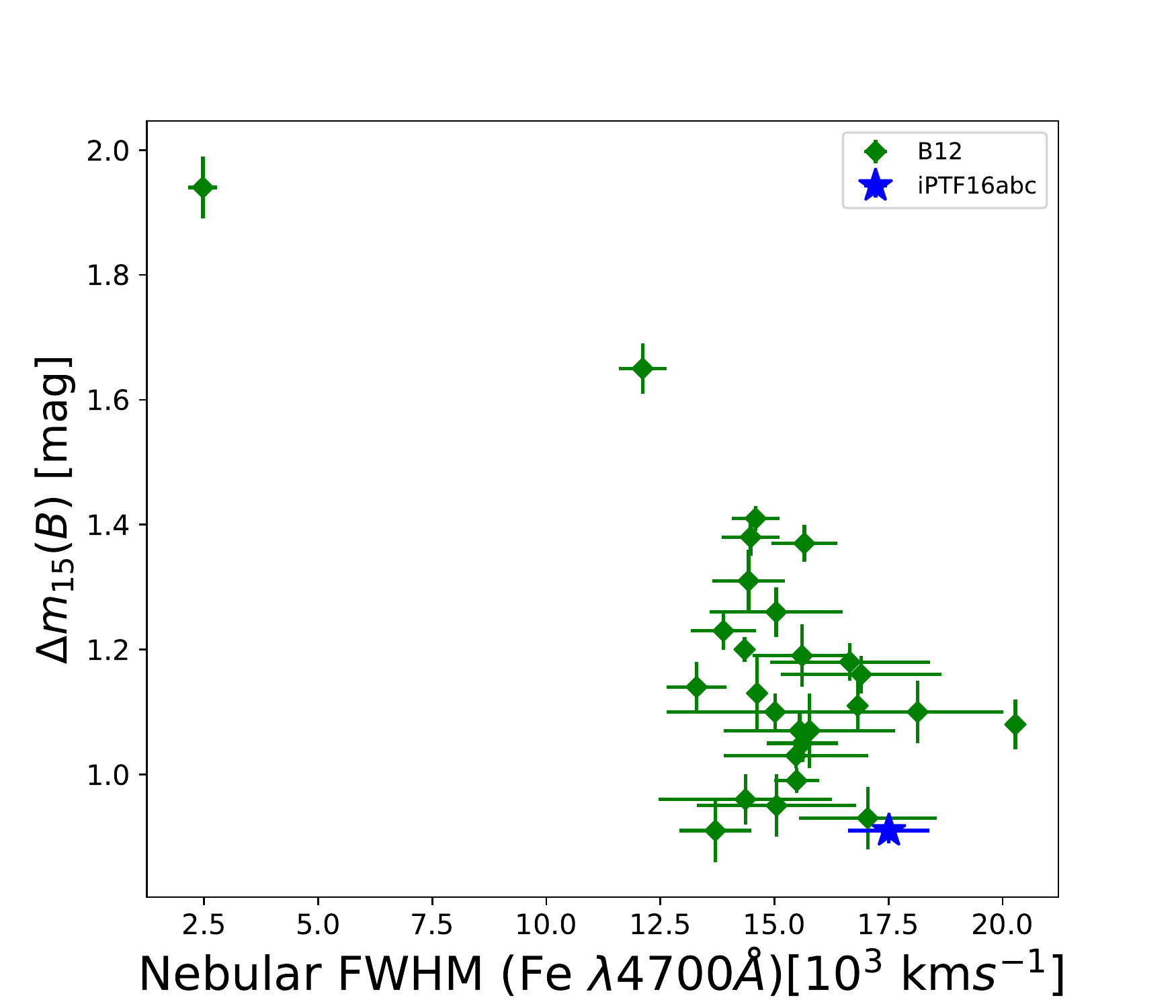}
\caption{The $\Delta m_{15} (B)$ for the supernovae from the CfA SN program \citep{Blondin2012} is plotted against the full width at half maximum (FWHM) of the Fe $\lambda$ 4700 \AA\, feature (green diamonds). iPTF16abc is plotted as the blue star and appears to be consistent with the distribution for the CfA sample.}
\label{fig:neb_dm15}
\end{figure}

We measured the full width at half maximum  of the iron feature 
at $\sim$ 4700\AA\, to be 17\,510 $\pm$ 891 km\,s$^{-1}$. \citet{Mazzali1998} found a relation between the FWHM of this feature and $\Delta m_{15}(B)$, although with a large sample of SNe \citet{Blondin2012} find no strong relation. In Figure~\ref{fig:neb_dm15}, we plot the FWHM of the 4700\AA\, feature for iPTF16abc with the sample of SNe from \citet{Blondin2012} and find that it is towards the higher end of the distribution of line widths but is consistent with the rest of the SNe in the literature.

\input{nir_table}
\subsection{Bolometric properties}
\label{ssec-bol}
Bolometric properties of SNe~Ia have been demonstrated to hold important information regarding the underlying physical parameters of the progenitors, e.g. \ejm\, , \Nif\, mass \citep{Contardo2000,Stritzinger2006a,Scalzo2014}. We compute the bolometric light curves for SNe with photometry from $u$ to $H$ filters. The observed magnitudes are dereddened using the reddening law in \citet{Cardelli1989}. Specifically, $E(B-V)$ and $R_V$ values are computed using \texttt{SNooPy} with the colour-stretch parameter \citep{Burns2011,Burns2014} and are summarized in Table~\ref{tab:ebv_rv}. Filter gaps and overlaps are treated as in \citet{Dhawan2016,Dhawan2017a} and the apparent flux is converted to absolute fluxes using distances in Table~\ref{tab:dist}. We use a cubic spline interpolation to derive the values of the maximum luminosity, \lm,~from the bolometric light curves. Here we compare the early and late time bolometric properties of a sample of Type Ia supernovae with the inferred values for iPTF16abc. Since the light curve shape for SNe~Ia can depend on the host galaxy reddening for highly-reddened SNe \citep{Leibundgut1988,Amanullah2011,Bulla2018}, we only use SNe with $E(B-V) < 0.5$ mag. $^{56}$Ni mass (\Mni)~values are derived from \lm~using Arnett's rule, i.e. assuming that the instantaneous rate of energy deposition equals the output flux at maximum \citep{Arnett1982}: 

\input{nickel_mass}

\begin{equation}
\label{eq:eni}
\epsilon_{Ni} (t_R, 1~M_{\odot})= \\\alpha\cdot(6.45  \cdot 10^{43} e^{-t_R/8.8} + 1.45  \cdot  10^{43} e^{-t_R/111.3})\,{\rm erg\,s^{-1}}, 
\end{equation}
where $t_R$ is the bolometric rise time and $\alpha$ is the parameter that accounts for deviations from Arnett's rule. These departures from $\alpha=1$ could be due to the $^{56}$Ni distribution, however, detailed model calculations of SNe~Ia find that $\alpha = 1$ with a scatter of 10-15$\%$ \citep{Blondin2013,Blondin2017}. The impact of this on further parameter estimation is discussed below.
Assuming Arnett's rule is obeyed and a rise time of $t_R=$19\,days with an error of 3\,days \citep{Stritzinger2006a, Dhawan2016}, we can derive a \Mni\, from the computed value of \lm:

\begin{equation}
\label{eq:lm-ni}
\frac{\mathrm{M_{^{56}Ni}}}{\mathrm{M_{\odot}}} = \frac{L_{max}}{2.0 (\pm 0.3) \times 10^{43}  \mathrm{erg\,s^{-1}}}~~. 
\end{equation} 
For iPTF16abc (see Table~\ref{tab:ni}), the small difference in the inferred \Mni\, from \citet{Miller2017} is due to the different methods adopted for calculating the distances. We discuss the impact of the rise time in Section~\ref{ssec-rise}. We note that some SNe in our sample have also been studied in detail in the literature \citep[e.g., SN2007on;][]{Ashall2018}. We find our inferred \Nif\, mass in good agreement with the values they report.

\input{tab_t0_16abc_withrise}

We also calculate the transparency timescale ($t_0$) for the sample of SNe, defined by \citet{Jeffrey1999} as a  parameter that governs the time-varying $\gamma$-ray optical depth behaviour of a supernova. We determined $t_0$ by fitting the radioactive decay energy deposition to the late time (forty to ninety\,days) bolometric light curve:
\begin{multline}
\label{eq:dep}
E_{{\rm dep}} =  E_{{\rm Ni}} + E_{{\rm Co~e^{+}}} + [1 - {\rm exp(-\tau_\gamma)}]E_{{\rm Co~\gamma}} \\[2mm]= \lambda_{{\rm Ni}}{{\rm N_{Ni0}}}~{\rm exp(-\lambda_{{\rm Ni}}t)Q_{{\rm Ni~\gamma}}} \\[2mm]+ \lambda_{{\rm Co}}{\rm N_{Ni0}} {\frac{\lambda_{{\rm Ni}}}{\lambda_{{\rm Ni}}-\lambda_{{\rm Co}}}}[{\rm exp(-\lambda_{{\rm Co}}t)- exp(-\lambda_{{\rm Ni}}t)}]\\[2mm] \times \{Q_{\rm Co~e^{+}} + Q_{{\rm Co~\gamma}}[1 - {\rm exp(- \tau_\gamma)}]   \},
\end{multline}
 where the factor (1-exp(-$\tau_\gamma$)) is replaced by 1 for $^{56}$Ni since complete trapping of $\gamma$-rays occurs at early times, when most of the light curve is powered by $^{56}$Ni.
where $\lambda_{Ni}$ and $\lambda_{Co}$ are the e-folding decay times of 8.8 days and 111.3 days for \Nif\, and \Cof\, respectively.  $Q_{\mathrm{Ni}\,\gamma}$ (1.75 MeV) is the energy release per \Nif\, $\rightarrow$ \Cof\, decay. $Q_{\mathrm{Co}\,\gamma}$ (3.61 MeV) and $Q_{\mathrm{Co}\,e^{+}}$ (0.12 MeV) are the \g-ray and positron energies, respectively, released per \Cof\, $\rightarrow$ \Fef\, decay \citep[see][]{Stritzinger2006a}.  Equation~\ref{eq:dep} is only applicable in the optically thin limit, when the thermalized photons can freely escape.

$\tau_{\gamma}$ is the  mean optical depth, calculated by integrating from the point of emission to the surface of the ejecta \citep[see][for a derivation of the expression]{Jeffrey1999}. It has a simple t$^{-2}$ dependence, given as,
\begin{equation}
\label{eq:tau}
\tau_{\gamma} = \frac{t_0^2}{t^2}.
\end{equation}
 where $t_0$ is the transparency timescale, which by construction in \citet{Jeffrey1999} is the epoch at which the optical depth is unity.

The value of the transparency timescale can be directly related to the total ejecta mass (\ejm) with the following equation \citep{Jeffrey1999,Stritzinger2006a,Dhawan2017a}
\begin{equation}
\label{eq:ejm}
M_{\rm ej} = 1.38 \cdot \left(\frac{1/3}{q}\right) \cdot \left( \frac{v_e}{3000\, \mathrm{kms^{-1}}}\right)^2 \cdot \left(\frac{t_0}{36.80\,days}\right)^2 \mathrm{M_{\odot}}.
\end{equation}
Equation\,\ref{eq:ejm} encapsulates the capture rate of $\gamma$-rays in an expanding spherical volume for a given distribution of the radioactive source.  We discuss the typical parameter values for canonical SN~Ia models to map $t_0$ to \ejm, however, we note that we do not use any calculations for \ejm in our analysis and only use the observed $t_0$ values.
$q$ is a qualitative description of the distribution of the material within the ejecta, with one third being a uniform distribution and higher values reflecting more centrally-concentrated \Nif ( more typical of subluminous 1991bg-like SNe \citep{Mazzali1997}). The e-folding velocity $v_e$ provides the scaling length for the expansion, which is  $\sim$ 3000 km\,$s^{-1}$ for Chandrasekhar-mass ($M_{\rm ch}$) explosions corresponding to typical brightness SNe~Ia and on average slightly lower for sub-\mch WDs.  For both \mch and sub-\mch models it ranges between $\sim$ 2700 - 3200 km\,$s^{-1}$.   For an ejecta with Y$_{\rm e}$ = 0.5, which is appropriate for SNe~Ia, the $\gamma$-ray opacity is a constant of 0.025 cm$^2$ g$^{-1}$ \citep{Swartz1995}.  Given the range of possible values for $q$ and $v_e$, we do not proceed to infer M$_{ej}$ for the SNe in our sample.

Because the UVOIR light curve is not truly bolometric, there is an implicit assumption that the flux redward of the $H$ band and bluewards of the $u$ band have very small contributions to the bolometric light curve. This assumption is supported by modelling that shows that the infrared catastrophe does not occur until $\sim$ 1 year post explosion or later, whereas typical values of $t_0$ are between 20 and 50 days and the line-blanketing opacity in the UV remains high \citep{Blondin2015, Fransson2015}.



\begin{figure*}
\includegraphics[width=.45\textwidth]{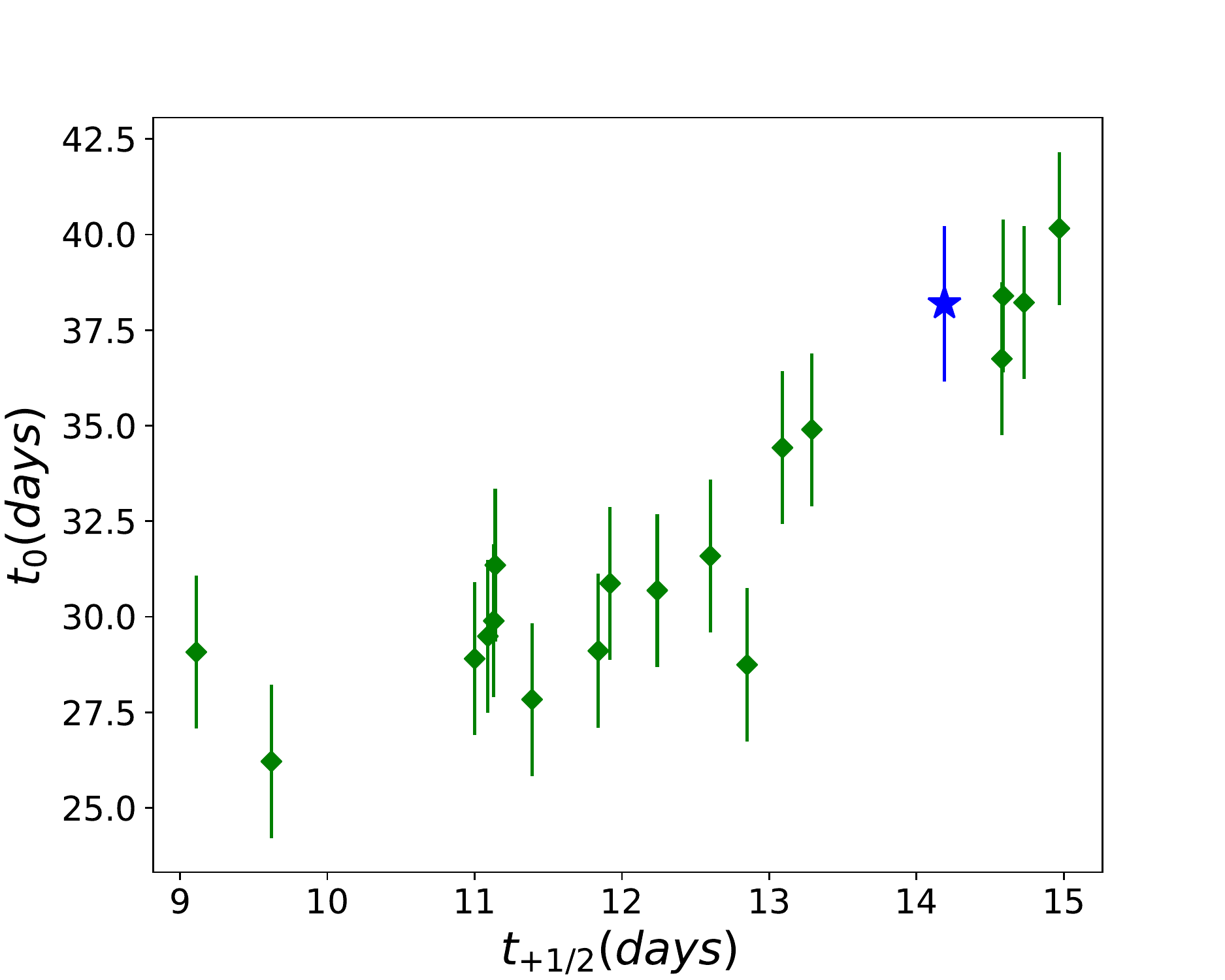}
\includegraphics[width=.45\textwidth]{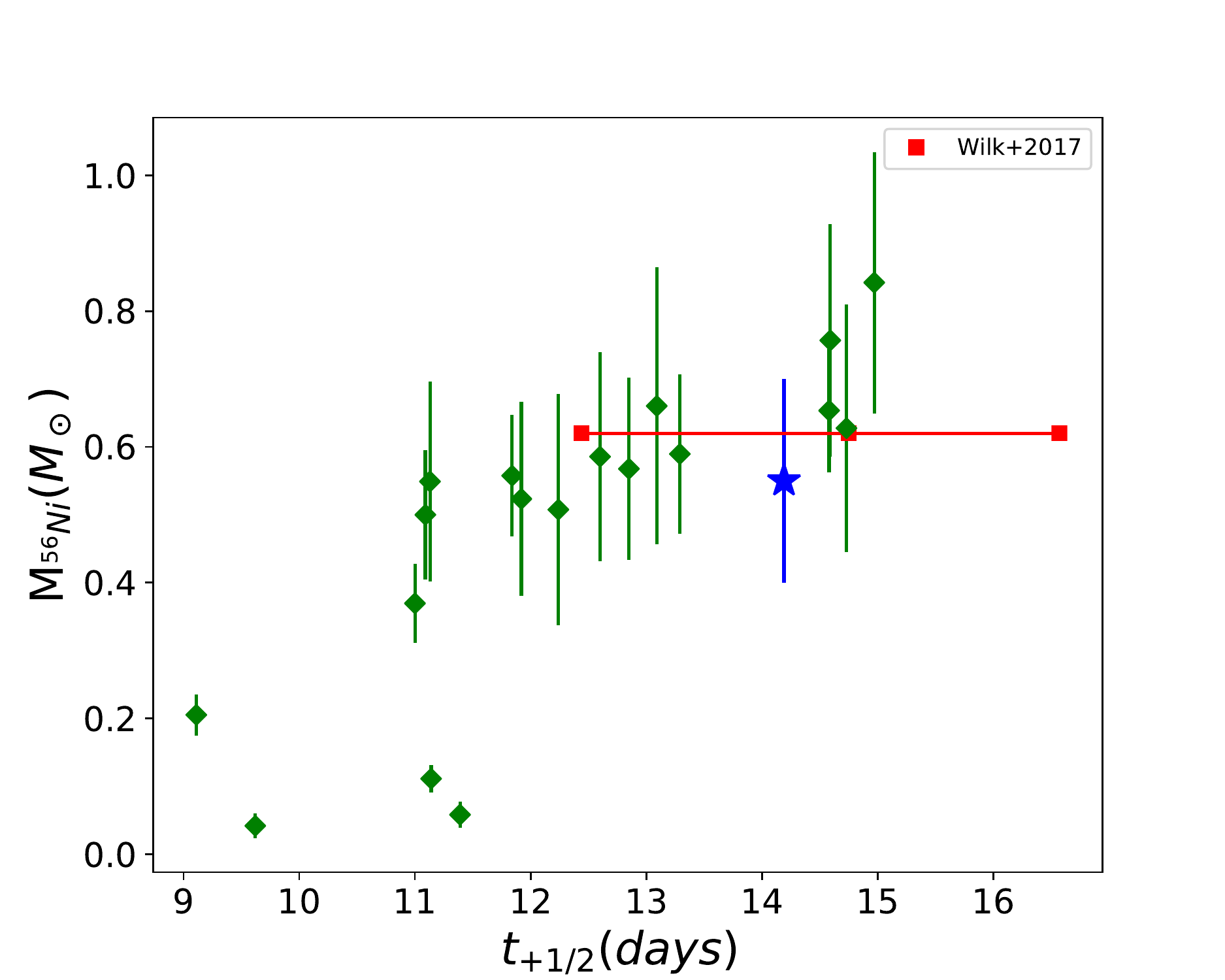}
\caption{(Left): $t_0$ is plotted against $t_{+1/2}$ for the sample of objects with sufficient late time data to measure $t_0$. There is a strong correlation between the two parameters, with $r \sim 0.85$. iPTF16abc is shown as the blue star and is consistent with the relation. (Right) \Mni\, is plotted against $t_{+1/2}$, the red points are predictions from the models of \citet{Wilk2017} for sub-M$_{Ch}$, M$_{Ch}$ and super-M$_{Ch}$ progenitors.}
\label{fig:thalf}
\end{figure*}

Recent studies of the diagnostics of ejecta mass \citep[e.g.][]{Wilk2017} compute the $t_{+1/2}$ for different ejecta mass models. $t_{+1/2}$ is defined as the time the bolometric light curve takes to decline to half of its peak luminosity \citep{Contardo2000}. In Figure~\ref{fig:thalf} we plot the $t_{+1/2}$ against $t_0$ for the sample with sufficient data. Unsurprisingly, we find a strong correlation ($r \sim 0.85$) for the objects in our sample, suggesting that the SNe that take longer to decline to half the peak luminosity also have optically thick ejecta for longer. Since \citet{Jeffrey1999} suggest that $t_0$ is directly related to \ejm and this equation has been used to calculate \ejm for SNe~Ia \citep{Stritzinger2006a, Scalzo2014}, this implies that $t_{+1/2}$ can also be used as a diagnostic for the ejecta mass.

\begin{figure}
\includegraphics[width=.5\textwidth]{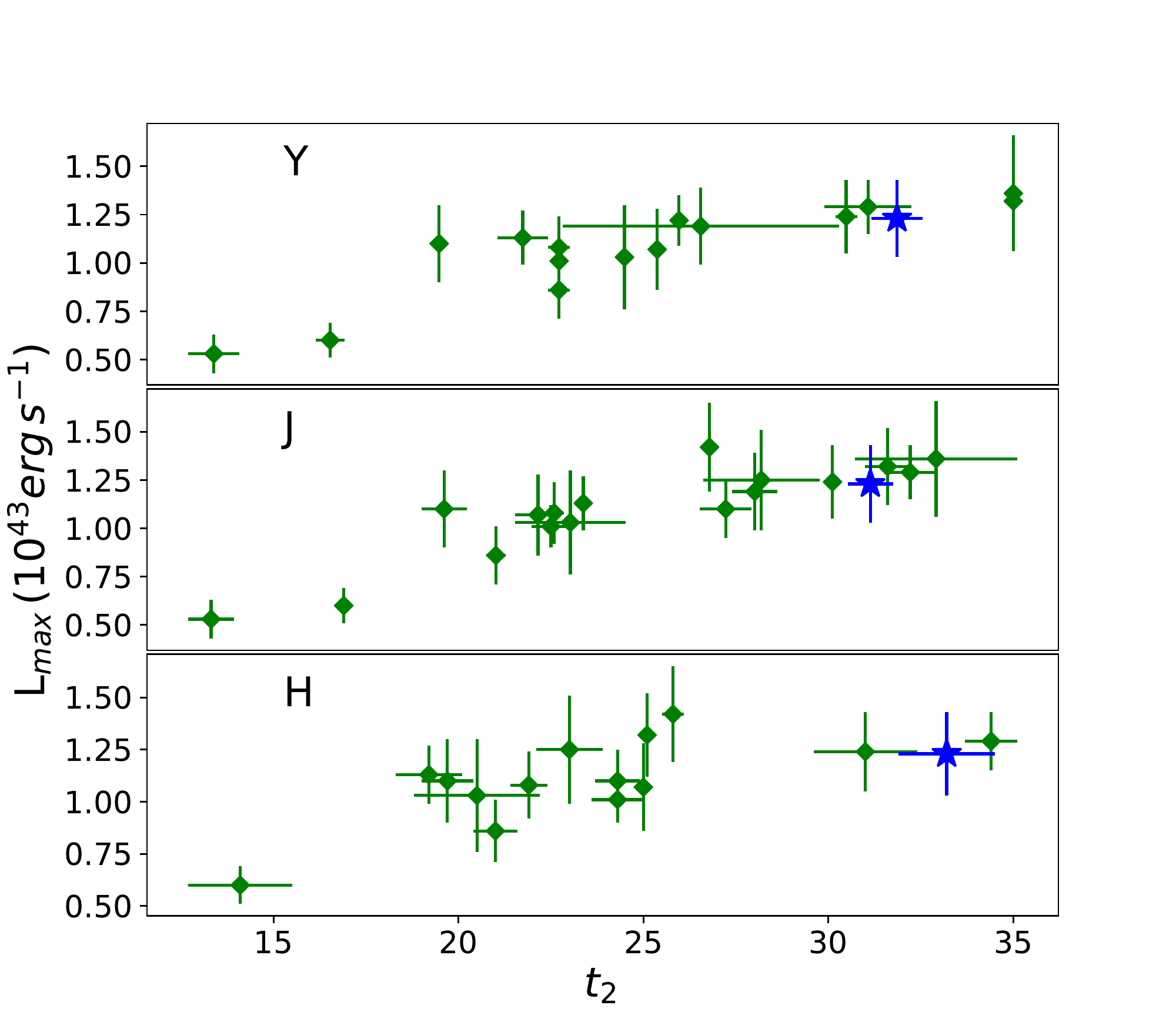}
\caption{The (pseudo-) bolometric peak luminosity is plotted against the timing of the second maximum in the $YJH$ filters. For both parameters, in this figure, the rise time is assumed to be 19\,days. The blue star is iPTF16abc. It lies on the relation for all other SNe~Ia, though towards the brighter side.}
\label{fig:lmax_t2}
\end{figure}

\subsection{Near Infrared light curves}
The near infrared morphology of SN~Ia light curves has been shown to be markedly different from the optical \citep{Elias1981,Elias1985}. The SNe show a characteristic second peak in the $izYJHK$ filters \citep{Mandel2009,Biscardi2012,Dhawan2015} and a less pronounced ``shoulder''-like feature in the $V$ and $r$ bands. 
Theoretical studies have shown that these features are caused by a recombination wave in the IGEs in the ejecta from doubly- to singly-ionised \citep{Kasen2006,Blondin2015}, hence, the NIR second maximum is an important parameter to test whether the observed peculiarities in iPTF16abc could be arising due to the properties of the central IGE core. 
iPTF16abc shows a distinct second maximum in the $YJH$ filters, consistent with the theoretical prediction for an SN~Ia that produced $\sim$ 0.6 $M_{\odot}$ of \Nif\, \citep{Kasen2006}. 

The values for the timing of the second maximum ($t_2$) for iPTF16abc are presented in Table~\ref{tab:nir} and shown in Figure~\ref{fig:lmax_t2}. The $t_2$ values of iPTF16abc are consistent within the range for normal SNe~Ia \citep{Biscardi2012,Dhawan2015}, although they are at the higher end  like 91T-like/99aa-like SNe. \citet{Dhawan2016} found a correlation between the peak (pseudo-) bolometric luminosity and $t_2$ for a sample of SNe with low reddening from the host galaxy dust. As shown in Figure~\ref{fig:lmax_t2}, the value for iPTF16abc lies on the expected relation. 

\input{photometry_de}

\subsection{Nebular Phase Photometry}
\label{ssec-neb_phot}
\begin{figure}
\includegraphics[width=.5\textwidth]{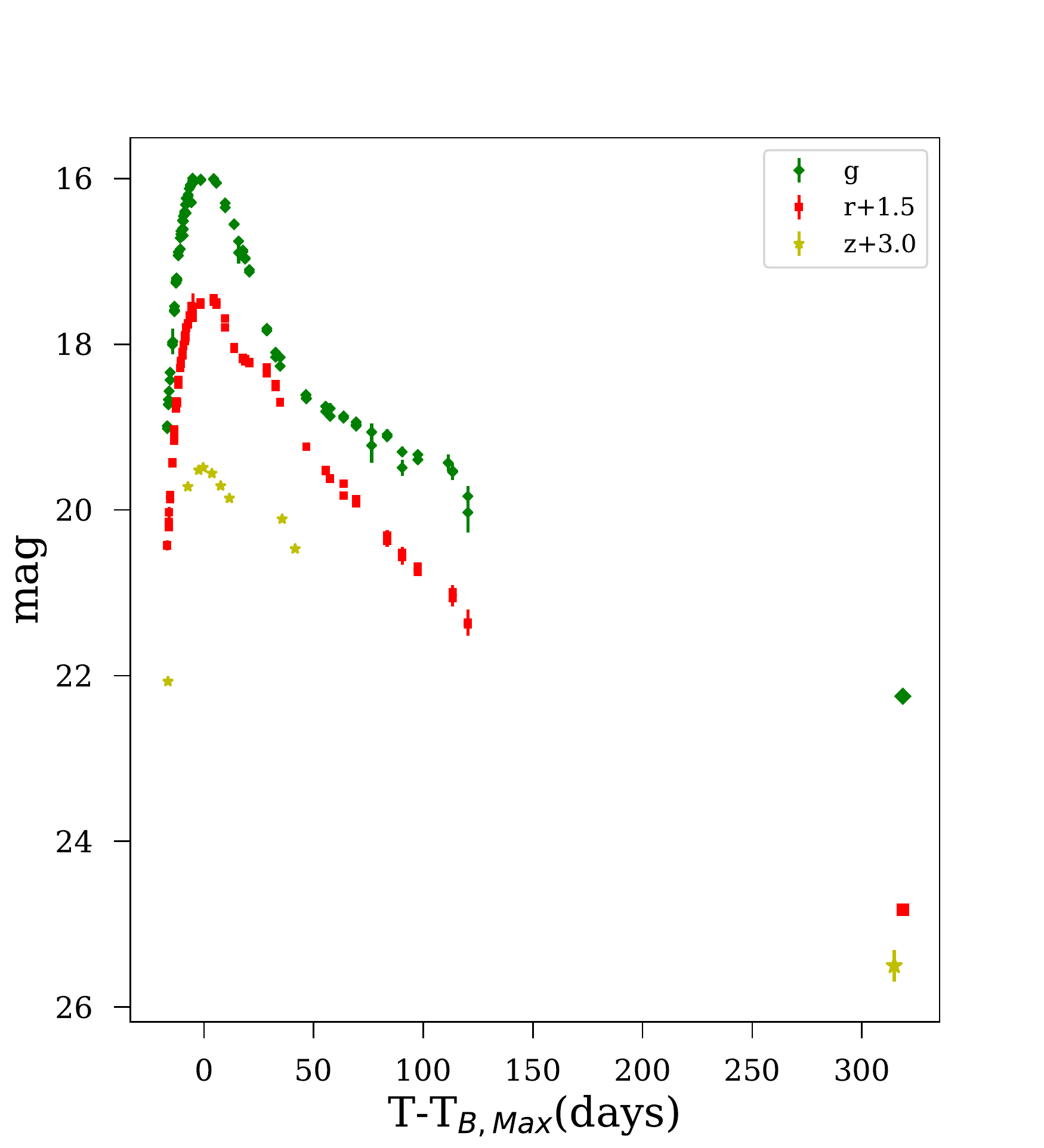}
\caption{$g$,$r$,$z$ light curves of iPTF16abc from \citet{Miller2017}, along with the nebular phase ($>$ + 300\,days) data points from the DECaLS survey. The legend gives the magnitude offsets for the $r$ and $z$ light curves.}
\label{fig:neb_lightcurve}
\end{figure}

iPTF16abc was also observed at epochs between +316-+320\,days by the DECaLS survey in $g,r,z$ filters. The photometry on the DECaLS images is presented in Table~\ref{tab:phot_decals}. The decrease of 6.6 mag from peak to $\sim$ +300\,days (see Figure~\ref{fig:neb_lightcurve}) is consistent with the expected value for normal SNe~Ia \citep[e.g.][]{Stritzinger2007}. We also find that the $r$-band absolute magnitude of -11.4 $\pm$ 0.11 mag is within the range of observed values in the literature \citep{Lair2006} at late phases of $\sim$ 1 year post maximum.  This further demonstrates that , at late epochs, the observed properties of iPTF16abc are within the observed distribution for normal SNe~Ia.  We note that since there is only $r$-band photometry available for SNe~Ia at these late epochs (out of the three filters presented here), it is the only one for which we can compare iPTF16abc to literature values for a sample.

\section{Discussion}
\label{sec-disc}
The early light curve and spectral evolution of iPTF16abc showed marked peculiarities that were interpreted as a signature of strong outward mixing of \Nif~and/or interaction with diffuse material \citep{Miller2017}. In the case of strong mixing, the temperature and thus ionisation level of the outer ejecta would be higher compared to the case of a centrally-concentrated distribution of \Nif~more commonly inferred for normal SNe~Ia. This would explain why in iPTF16abc not only C\,{\sc ii} features are stronger \citep{Miller2017} but also Si\,{\sc ii} and Ca\,{\sc ii} features are weaker compared to what is seen for the bulk of normal SNe~Ia. Shallow silicon and calcium features are also observed in peculiar events such as 2002cx-like and 1991T-like SNe~Ia. Models for both 2002cx-like  \citep[pure deflagration models, e.g.][]{Sahu2008,Kromer2013} and 1991T-like  \citep{Ruizlapuente1992,Mazzali1995,Sasdelli2014,Fisher2015,Seitenzahl2016,Zhang2016} SNe~Ia invoke the presence of \Nif~and IGEs in the outer ejecta, thus corroborating the idea that mixing might be responsible for the photometric and spectroscopic features seen in iPTF16abc. The nebular spectrum of iPTF16abc, however, looks consistent with the normal SN~2011fe, and the measured line width of the 4700 \AA\, agrees with the width-decline rate relation for a sample of normal SNe from the literature (Figure~\ref{fig:neb_dm15}). The likely source of these peculiarities could therefore arise from the outer layers of the ejecta, and probably not from the central core of IGEs. 

\begin{figure}
\includegraphics[width=.5\textwidth]{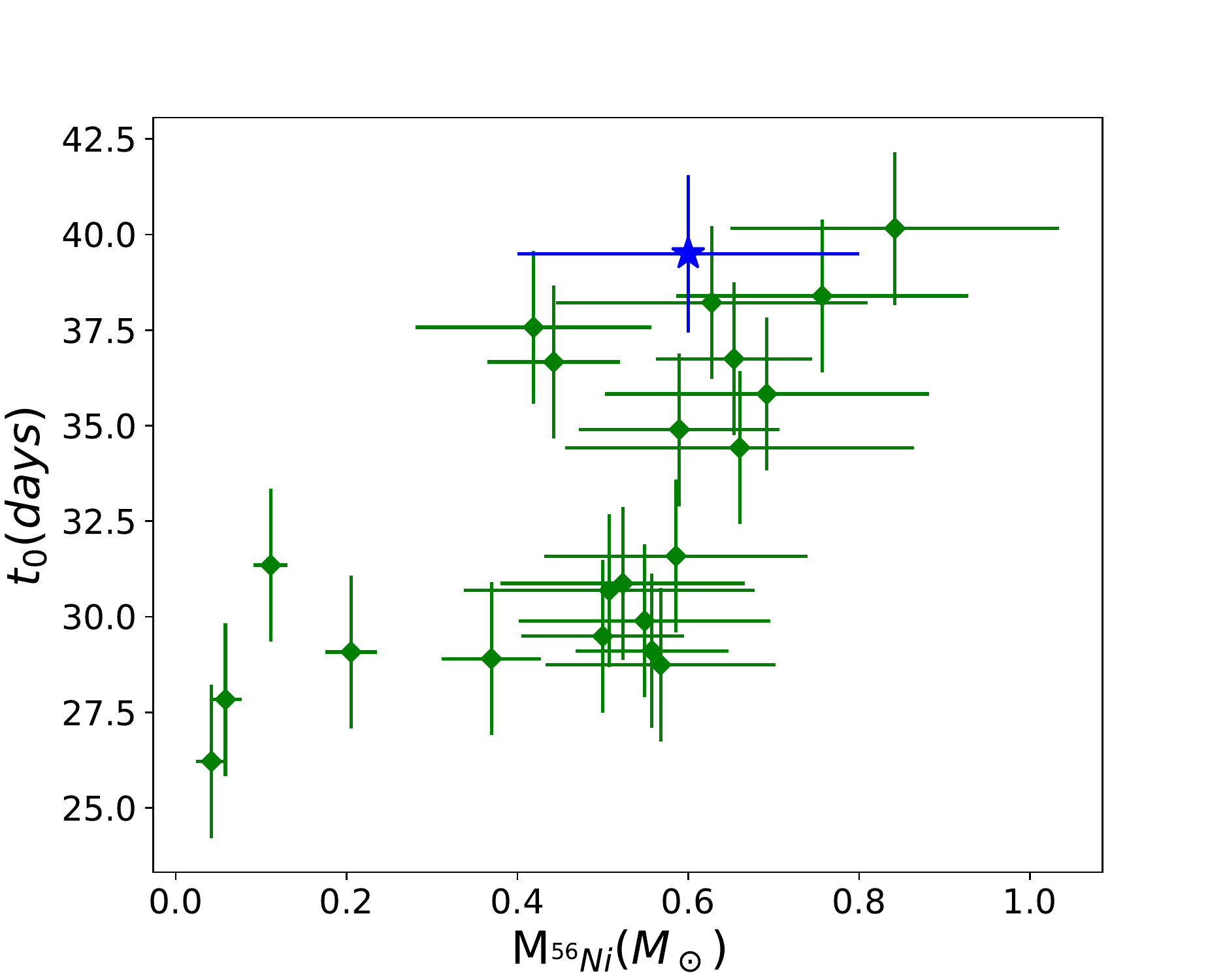}
\caption{The transparency timescale, $t_0$, is plotted against the $^{56}$Ni mass for a sample of SN~Ia with $u \rightarrow H$ band coverage (green diamonds). There is a weak correlation ($r \sim 0.67$) between the two parameters. iPTF16abc is plotted as a blue star and appears to be consistent with the general trend of $t_0$ versus $M_{^{56}\rm Ni}$.}
\label{fig:global}
\end{figure}

The transparency timescale is strongly tied to the ejecta mass of the SN \citep{Jeffrey1999,Stritzinger2006a,Scalzo2014},  therefore, given the uncertainties in driving the ejecta masses from $t_0$, we use $t_0$ directly for our analysis. As shown in Figure~\ref{fig:global}, we find that the total \Mni\, is correlated with the transparency timescale for our sample of SNe~Ia such that objects with a longer $t_0$ produce more \Mni\, also indicating a possible positive relation of the ejecta mass with the amount of synthesized \Nif\, in agreement with previous studies \citep{Stritzinger2006a,Scalzo2014}. iPTF16abc is consistent with this relation between $t_0$ and \Mni. For both parameters, we assume a canonical rise time of 19\,days. In Section~\ref{ssec-rise}, we discuss the impact of this assumption. 

\input{thalf_vals}
The right panel of Figure~\ref{fig:thalf} shows a weak correlation between the half light time for the bolometric light curve ($t_{+1/2}$) and the \Mni\, as in \citet{Contardo2000}. Comparing the observed $t_{+1/2}$ from models for three different ejecta masses \citep{Wilk2017}, all producing 0.6 $M_{\odot}$ of \Mni~\citep[i.e. the same as the estimate for iPTF16abc][]{Miller2017} we find that the range of $t_{+1/2}$ observed in SNe~Ia is higher, possibly indicating that there is a large range of ejecta masses to explain the observations of the sample of SNe~Ia. The observed $t_{+1/2}$ range could indicate that the sample of SNe corresponds to a range of ejecta masses. Since the lowest \ejm investigated in \citet{Wilk2017} is 1.02 $M_{\odot}$, lower $t_{+1/2}$ could imply smaller ejecta masses for the lower luminosity SNe in the sample \citep[see also][]{Blondin2017}.

\subsection{Rise time}
\label{ssec-rise}

In this section we discuss the effect of using different prescriptions for the rise time as an input to calculate the \Mni\, and $t_0$. Although iPTF16abc has early time observations that can strongly constrain the rise \citep[$\sim$ 18\,days in the $g$-band,][]{Miller2017}, for a sample of SNe, the early behaviour is not as well-characterised. Our canonical assumption in Section~\ref{sec-res} was to use a rise time of 19\,days along with an error of 3\,days to capture the diversity in the observed values for the population of SNe~Ia \citep[e.g.][]{G11}. In \citet{G11}, the authors find that the rise time is correlated with the post-peak decline rate and they derive a linear relation between the $B$-band rise time and $\Delta m_{15,B}$: 
\begin{equation}
t_{R, B}=17.5 - 5 \cdot (\Delta m_{15, B} - 1.1)
\end{equation}
The bolometric maximum occurs on average one~day before $B_{max}$ \citep{Contardo2000,Scalzo2014}, hence, we derive the rise time using
\begin{equation}
\label{eq:bol_rise}
t_{R, bol}=16.5 - 5 \cdot (\Delta m_{15, B} - 1.1)
\end{equation}

Hence, faster declining SNe~Ia would have a faster rise. From equation~\ref{eq:eni} we can see that a faster rise time implies a smaller \Mni\, for the same \lm. The error on the rise from this method is $\sim$ 2 days \citep{Scalzo2014}. The impact of the rise time on the inferred \Mni\, is shown in Table~\ref{tab:ni}. It is clear from the table that even with a varying rise time across the sample, the \Mni\, estimates are within the errors.

\begin{figure}
\includegraphics[width=.5\textwidth]{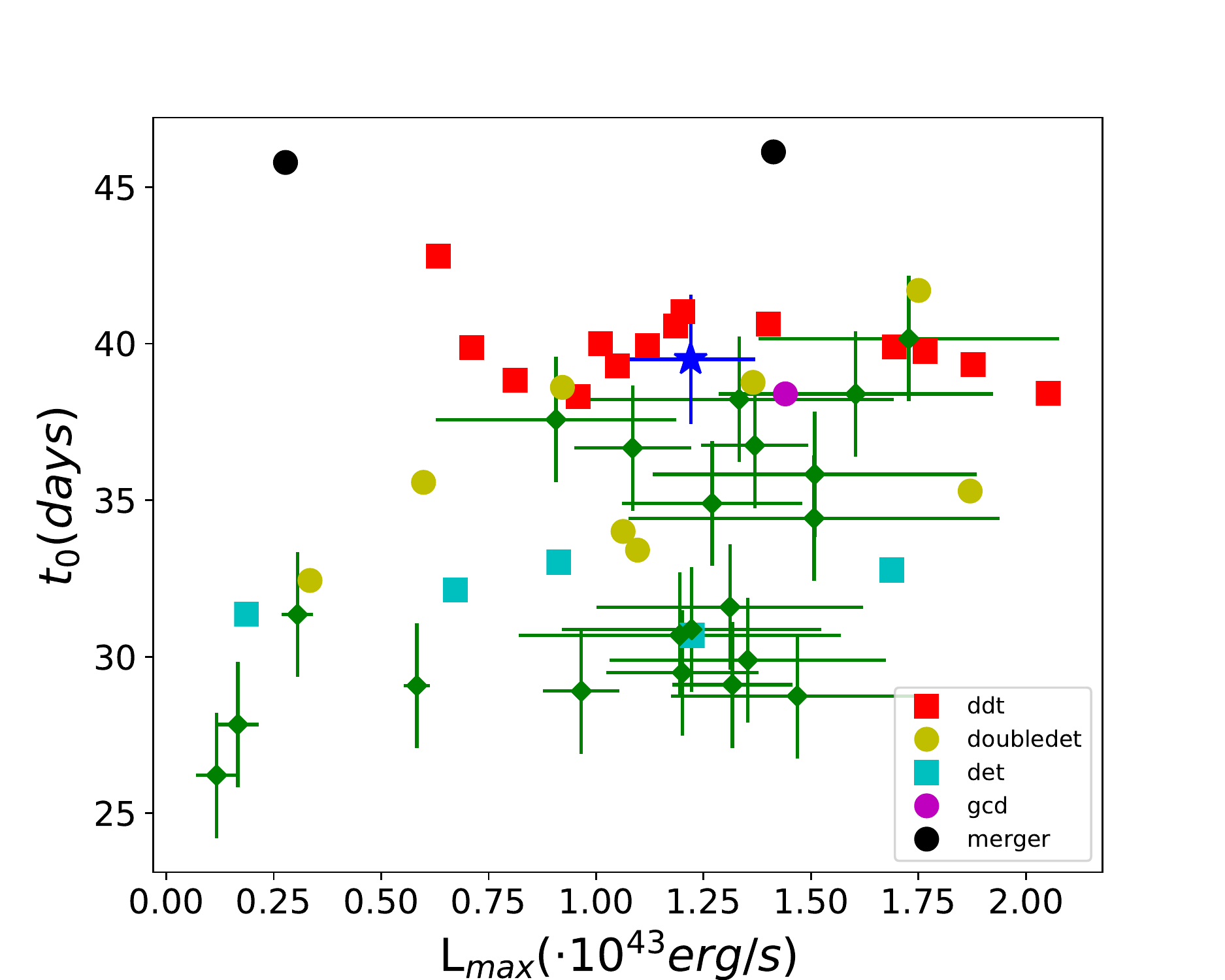}
\caption{The transparency timescale is plotted against the peak bolometric luminosity for the sample of SNe in this study. Overplotted are the expectations from the different explosion scenarios. The red squares are the three dimensional \mch delayed detonation models of \citet{Seitenzahl2013}, the magenta circle is the \mch gravitationally confined detonation of \citet{Seitenzahl2016}. The yellow circles are the sub-\mch ``double-detonation'' models of \citet{Fink2010} whereas the yellow squares are the sub-\mch pure detonations presented in \citet{Sim2010}. The two black circles are the violent merger models of \citet{Pakmor2010} and \citet{Pakmor2012}. }
\label{fig:ejecta_nickel}
\end{figure}
We note that the estimation of $t_0$ from the (pseudo)-bolometric light curve, depends on the inference of the number of \Nif\, atoms, which in turn is sensitive to the 
$\alpha$ parameter. Since the deviations from the $\alpha=1$ case are of the order 10-15$\%$, the difference in the estimated $t_0$ is within the errors. 

\subsection{Comparison with explosion models}
\label{ssec-model}
We computed $t_0$ and \Mni\, for a sample of SNe and found that these quantities were weakly correlated \citep[see also][]{Stritzinger2006a,Scalzo2014}. iPTF16abc lies on the $t_0$ - \Mni\, relation for the SNe studied here, suggesting that the amount of \Mni\, and the progenitor mass for the SNe is compatible with the normal population of SNe~Ia. Here, we compare the relation between $t_0$ and \lm\, to predictions from different theoretical models, since we want to directly compare observable quantities from the data to the model predictions. A range of $t_0$ values from $\sim$ 25 to 40\,days in itself would suggest that not all SNe~Ia could arise from the same ejecta mass \citep{Stritzinger2006a,Scalzo2014}. 

In Figure~\ref{fig:ejecta_nickel} we plot the $t_0$ against the \lm\, for the sample of SNe in this study and compare them to the predictions from various explosion scenarios. The synthetic observables were taken from the Heidelberg Supernova Model Archive \citep[HESMA;][]{Kromer2017}. We find that the bright end of the distribution appears to be consistent with the \mch delayed detonation models of \citet{Seitenzahl2013}. The faint end of the relation is traced well by the sub-\mch models of \citet{Fink2010} and \citet{Sim2010} although some of the predictions from the most massive sub-\mch models seem consistent with the brightest observed SNe too. The violent merger models imply a longer transparency timescale than observed for any SNe in the sample. iPTF16abc is consistent with the observed relation and appears to be closer to the trend for the \mch models. We find our conclusion that sub-\mch scenario are consistent with nearly the entire range of observed $L_{max}$ and $t_0$ whereas the \mch can only explain the bright end is supported by recent theoretical studies which find that the entire width-luminosity relation reproducable by sub-Chandra models whereas the Chandrasekhar mass models only reproduce the bright end \citep{Goldstein2018}. This conclusion is  also supported by a comparison of observations and model predictions presented in \citet{Shen2017}.
 The two overluminous SNe in our sample with \texttt{SNID} classifications of 91T-like \citep{Folatelli2013} are both consistent with M$_{ch}$ model prediction.  \citet{Fisher2015} suggest that M$_{ch}$ WD explosions that lack a vigourous deflagration phase would produce 91T-like SNe. Since these explosions would also have higher \Nif\, mixing, this might indicate that overluminous SNe that show 91T-like spectral features have shorter deflagration phases.

We note that iPTF16abc appears to have $t_0$ estimates consistent with $M_{\rm ej} = 1.4 M_{\odot}$, under the assumption that $q \sim 1/3$. This value of $q$ would point towards strong \Nif\, mixing in the ejecta  in line with the findings of \citet{Miller2017}. \citet{Miller2017} cite a discrepancy between the early time observables and the early colours predicted for the sub-Chandrasekhar ``double-detonations'' of \citet{Noebauer2017} as a reason that iPTF16abc is likely not a result of the double detonation of a sub-\mch WD. The late-time bolometric properties would point to a similar conclusion.

Using the transparency timescale we can calculate the epoch at which the energy deposition from the $\gamma$ rays is equal to the contribution from the positrons \citep[hereafter, $t_c$, see][]{Childress2015}. For iPTF16abc, $t_c = 205.2 \pm 11.3$\,days which is consistent with the calculated value for the \mch delayed detonation model (DDC10) of \citet{Blondin2013} of $t_c \sim 214$ \,days, compared to values between 160 and 180\,days for sub-\mch models \citep{Sim2010,Fink2010}. This direct comparison adds further evidence for iPTF16abc being the result of \mch explosion. 

\subsection{Comparison to overluminous SNe~Ia}
In figure~\ref{fig:ejecta_nickel} we see that brighter objects have longer transparency timescales and these objects are more consistent with the predictions from Chandrasekhar mass models than sub-Chandra models. Some of these high luminosity SNe are classified as 91T-like \citep[e.g. SN2005M, SN2007S][]{Folatelli2013} whereas others are classified as normal \citep[e.g. SN2006ax, SN2008bc][]{Folatelli2013}. Hence, this would suggest that overluminous SN~Ia have similar maximum and post-maximum bolometric properties and possibly similar progenitor masses, independent of the early time spectroscopic differences \citep[e.g.][]{Miller2017,Contreras2018}. Overluminous SNe with early  observations also show a linear rise ($\sim$ 3 days after first light) which could be signs of strong mixing of \Nif\, \citep{Magee2018}. Hence the extent of \Nif\, mixing could be a parameter governing the early time photometric and spectroscopic diversity of SNe~Ia. These distinct spectroscopic features are also seen in M$_{ch}$ models which lack a vigorous deflagration phase \citep{Fisher2015}, further adding to the evidence that the progenitors of overluminous SNe~Ia could be Chandrasekhar mass.

The  width of the [Fe\,III] 4700 \AA line in the nebular phase spectrum for iPTF16abc is consistent with the values for normal SNe~Ia, but towards the higher end of the distribution. This might indicate that the extent of the IGE core is higher than average for SNe~Ia, possible further evidence for strong mixing in the ejecta. It is, however, hard to draw strong conclusions since the nebular spectra in the optical consist of several line blends.

\section{Conclusions}
\label{sec-conc}
iPTF16abc has an excellent set of early time observations that showed a peculiar rise time and unusual spectroscopic features \citep{Miller2017}. In this study, we present a nebular spectrum of the SN and nebular phase photometry in the optical. We analyse the early and late time observations of iPTF16abc in context of a sample of SNe~Ia from the literature. 

We measure the bolometric peak luminosity, \Nif\, masses and transparency timescales for 21 SNe and find a weak correlation between $t_0$ and \Nif\, mass indicating that SNe that produce more \Nif\, also become transparent at a later epoch \citep[see also][]{Stritzinger2006a,Scalzo2014}. iPTF16abc lies on the relation between $t_0$ and \Mni, although it is on the brighter end of the distribution. Additionally, we measure $t_{+1/2}$, the bolometric time of half-light and compare the inferred values to the predictions from models of \citet{Wilk2017}. The observed range of $t_{+1/2}$ values is larger than the range from the models which have ejecta mass values between 1.02 and 1.70 $M_{\odot}$, implying a large range of ejecta masses for the SNe. 

The  transitional and  nebular spectrum of iPTF16abc appear qualitatively similar to the normal SN~2011fe  as well as to overluminous SNe 1991T and 1999aa. The absolute magnitude of iPTF16abc in the nebular phase is consistent with normal SNe from the literature. iPTF16abc is also consistent with the relation between the FWHM of the 4700\AA\, feature and $\Delta m_{15}(B)$ for SNe from the literature \citep{Blondin2012,Silverman2013}, though it lies towards the high line width end of the relation. This would suggest that the peculiarities seen in the early time spectra of iPTF16abc are not present at late epochs, hence, arguing for obtaining data at both early and late epochs to understand the explosion properties of SNe~Ia.  

From Figure~\ref{fig:ejecta_nickel}, we find that the sub-\mch models appear to explain a large fraction of the $t_0$ - \lm\, relation, though the bright end is still better explained by the \mch delayed detonation. This could be indicative of multiple explosion mechanisms explaining the observed diversity of SNe~Ia. iPTF16abc appears to be consistent with the properties of \mch explosion scenario.

\section*{Acknowledgements}
We would like to thank Peter Nugent for pointing us to the DECaLS photometry of iPTF16abc. We acknowledge fruitful comments from Jesper Sollerman.  Funding from the Swedish Research Council, the Swedish Space Board and the K\&A Wallenberg foundation made this research possible. A.A.M. is funded by the Large Synoptic Survey Telescope Corporation in support of the Data Science Fellowship Program.  This work made use of the Heidelberg Supernova Model Archive (HESMA), https://hesma.h-its.org.






\onecolumn
\appendix
\section{Radioactive Decay Energy fits to bolometric light curves}
\label{sec:appendix}
In Section~\ref{ssec-bol}, we describe the procedure for fitting a radioactive decay energy deposition function to the bolometric light curve. Here, we present the light curve fits for the SNe in our sample. We plot the deposition curve for the best fit value of the transparency timescale and along with it the curve for complete trapping and complete escape. 

\begin{figure}
\centering
\includegraphics[width=.2\textwidth]{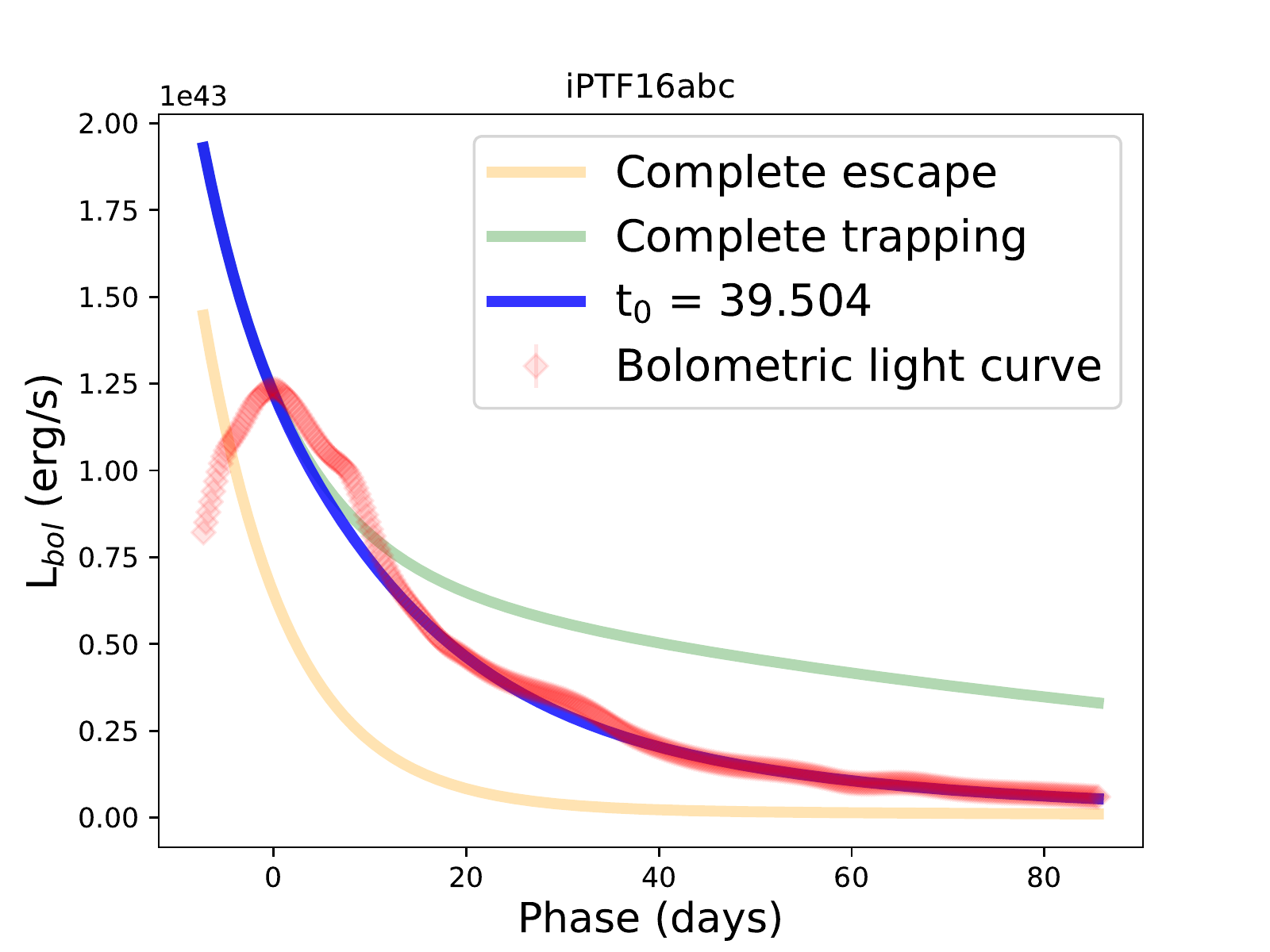}
\includegraphics[width=.2\textwidth]{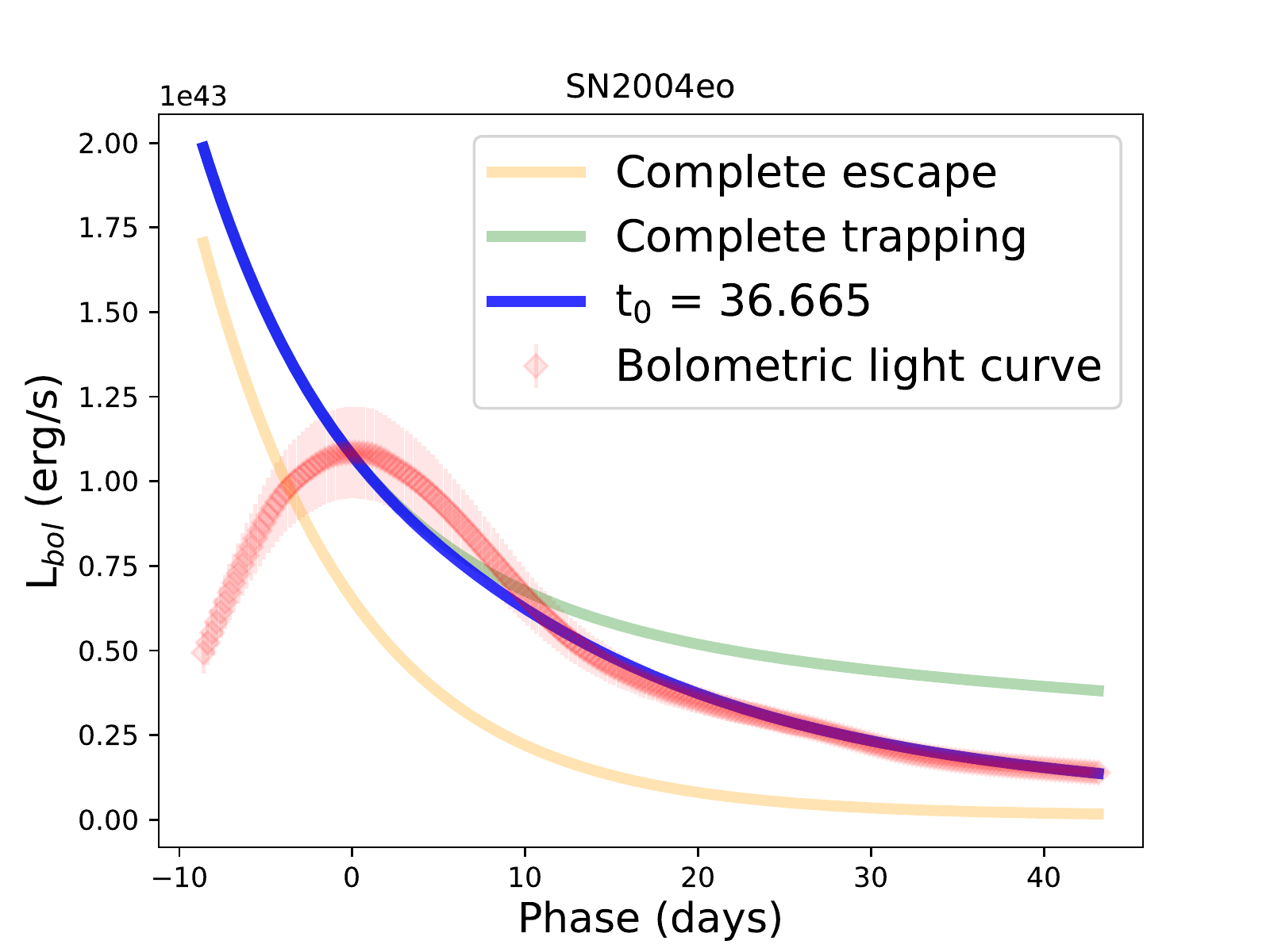}
\includegraphics[width=.2\textwidth]{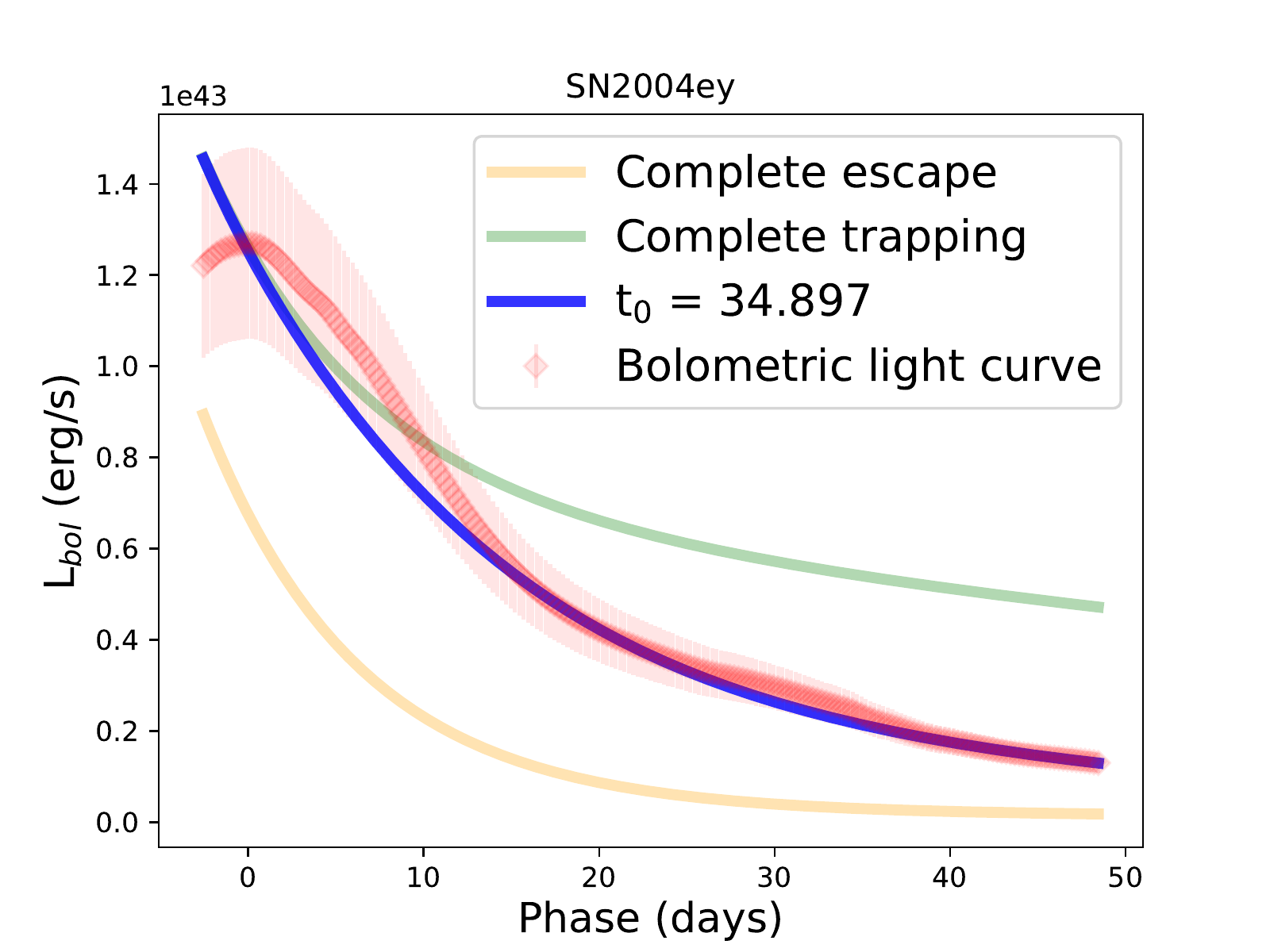}
\includegraphics[width=.2\textwidth]{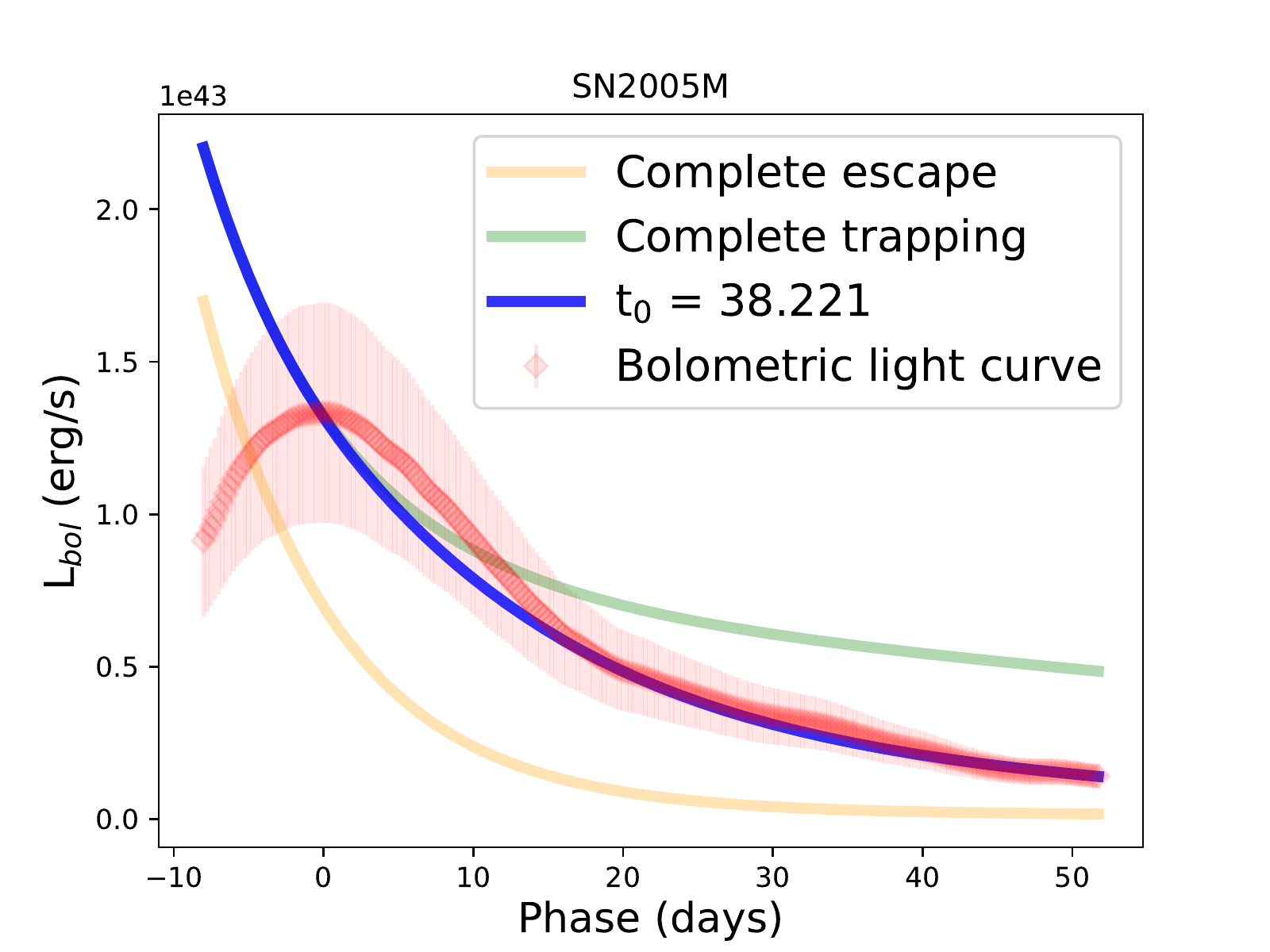}
\includegraphics[width=.2\textwidth]{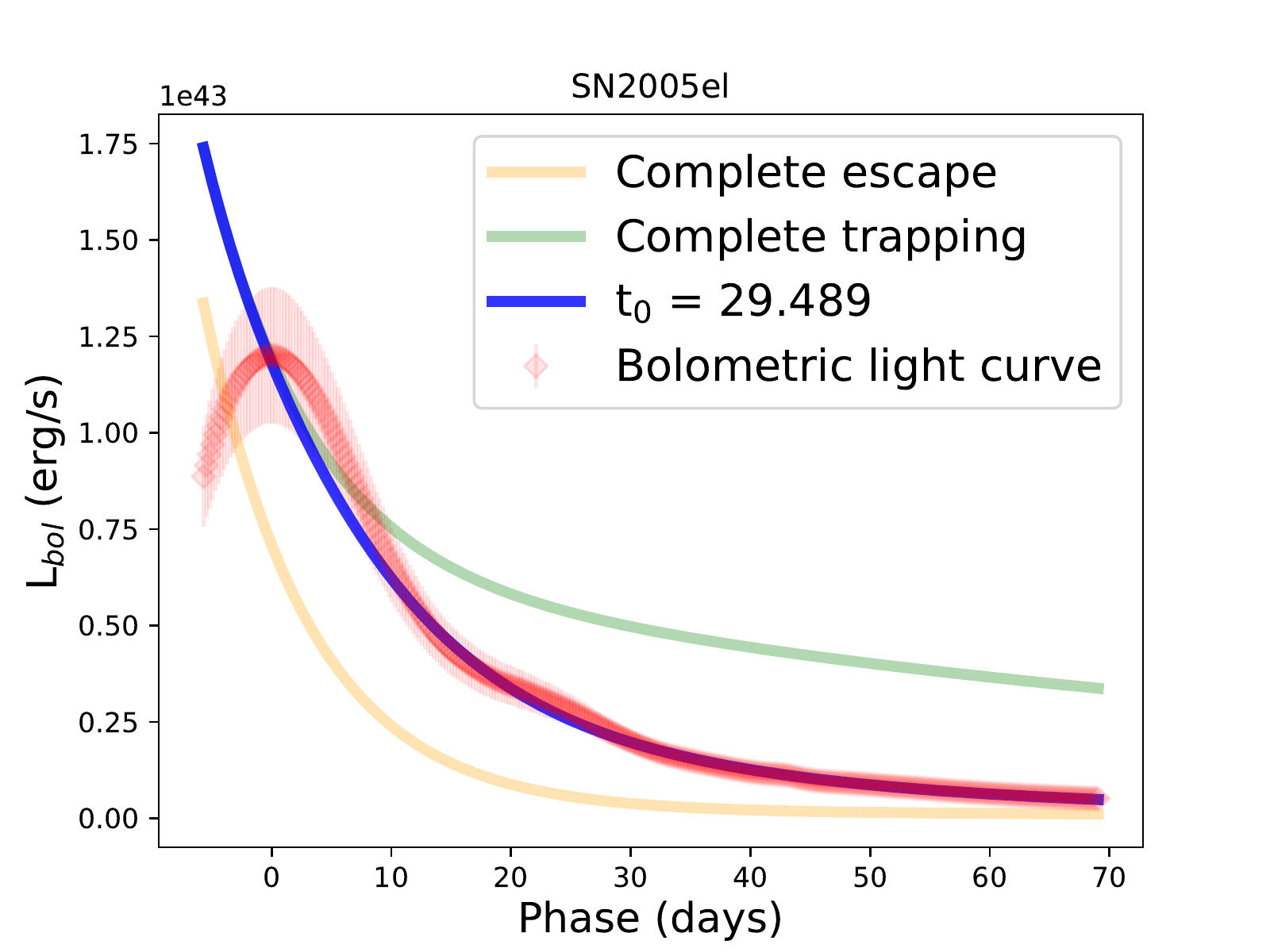}
\includegraphics[width=.2\textwidth]{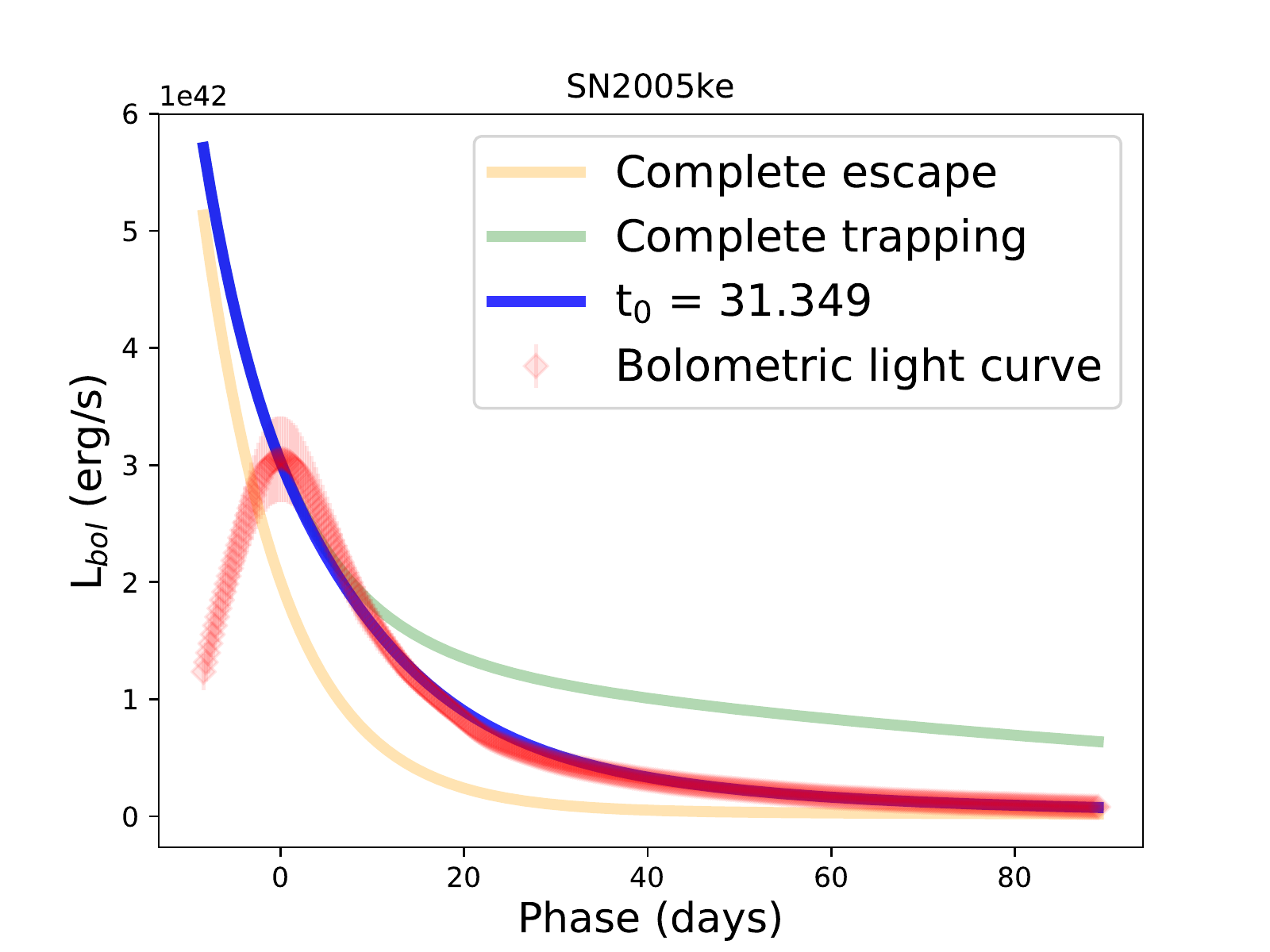}
\includegraphics[width=.2\textwidth]{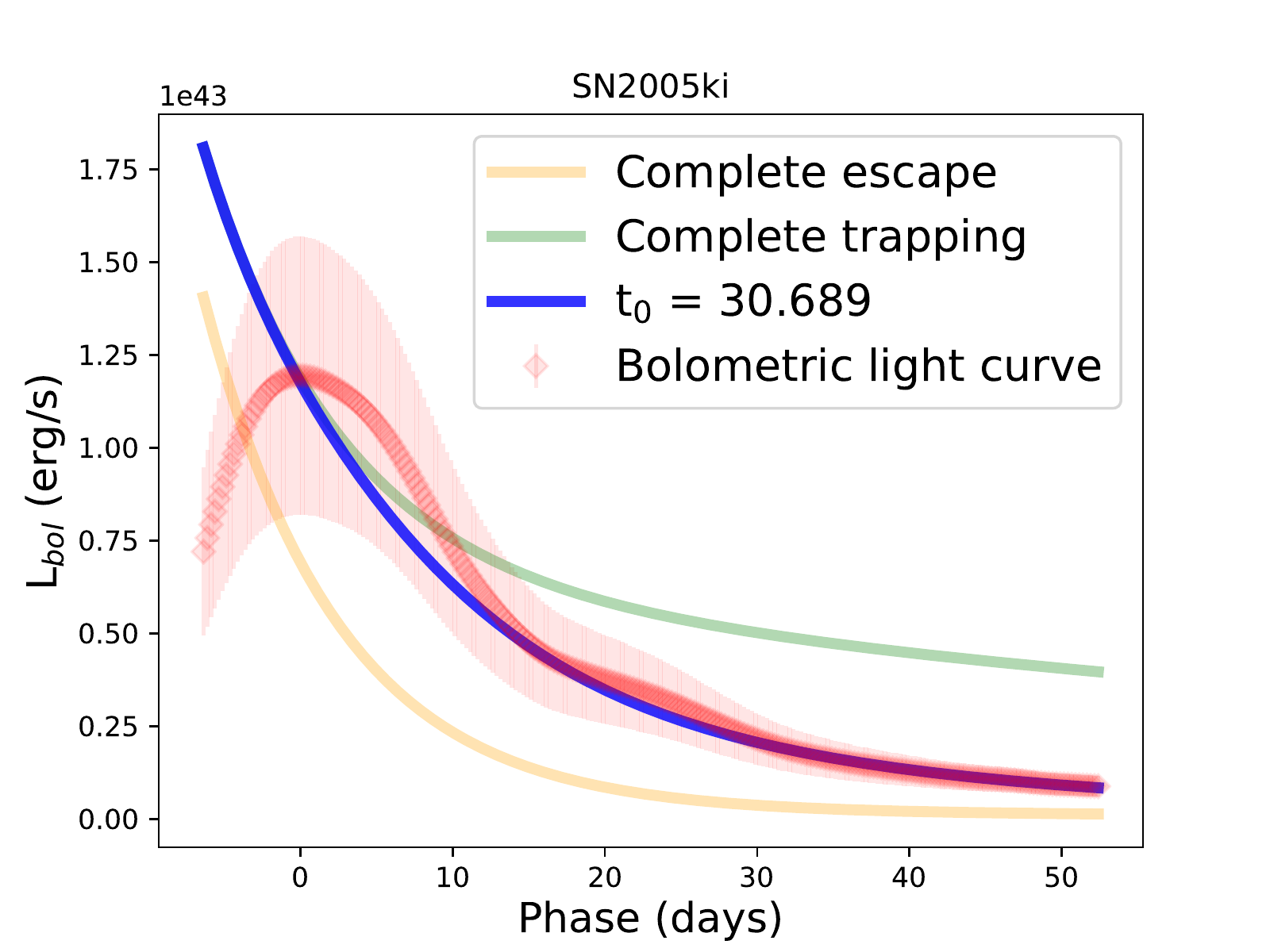}
\includegraphics[width=.2\textwidth]{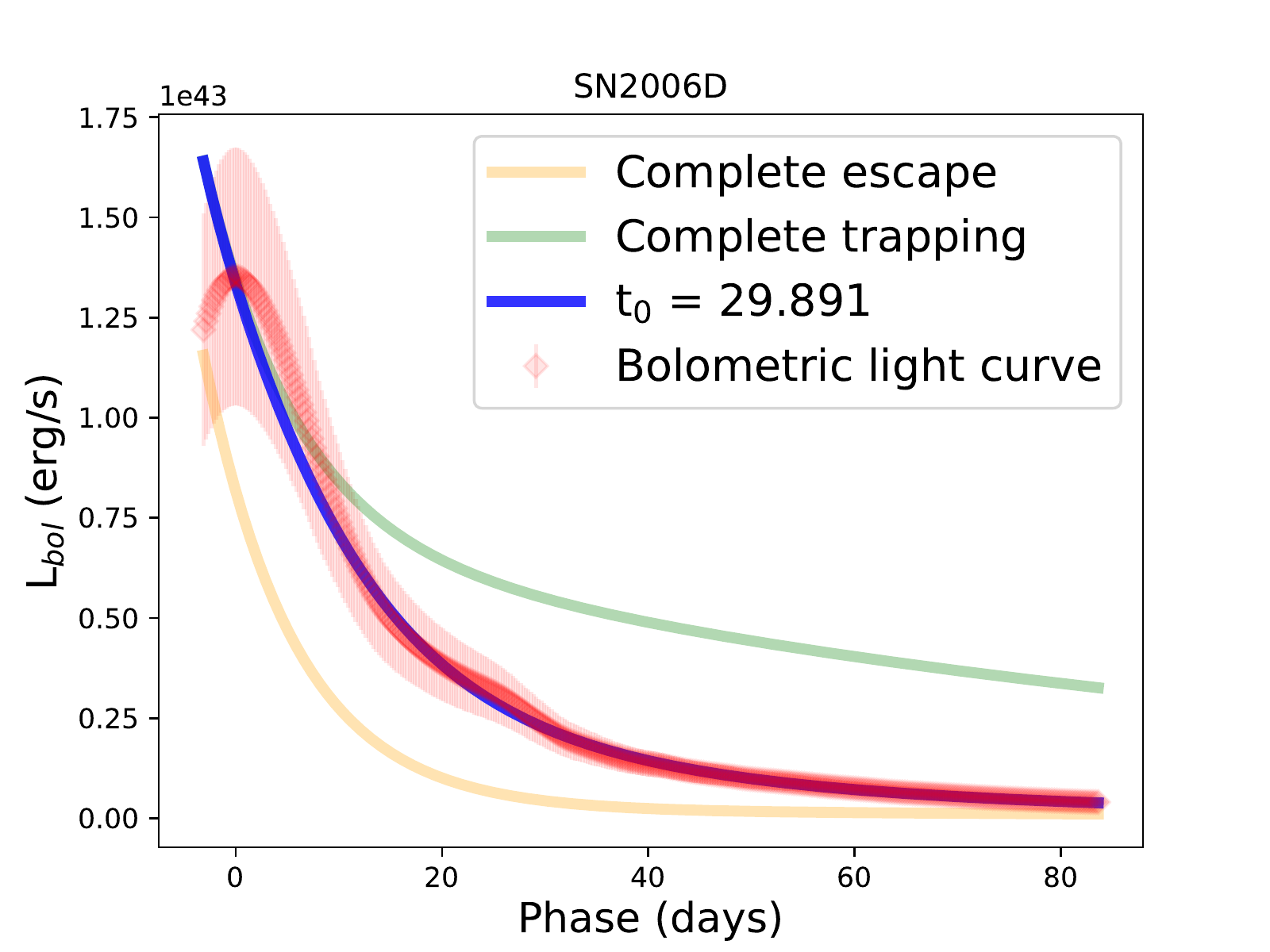}
\includegraphics[width=.2\textwidth]{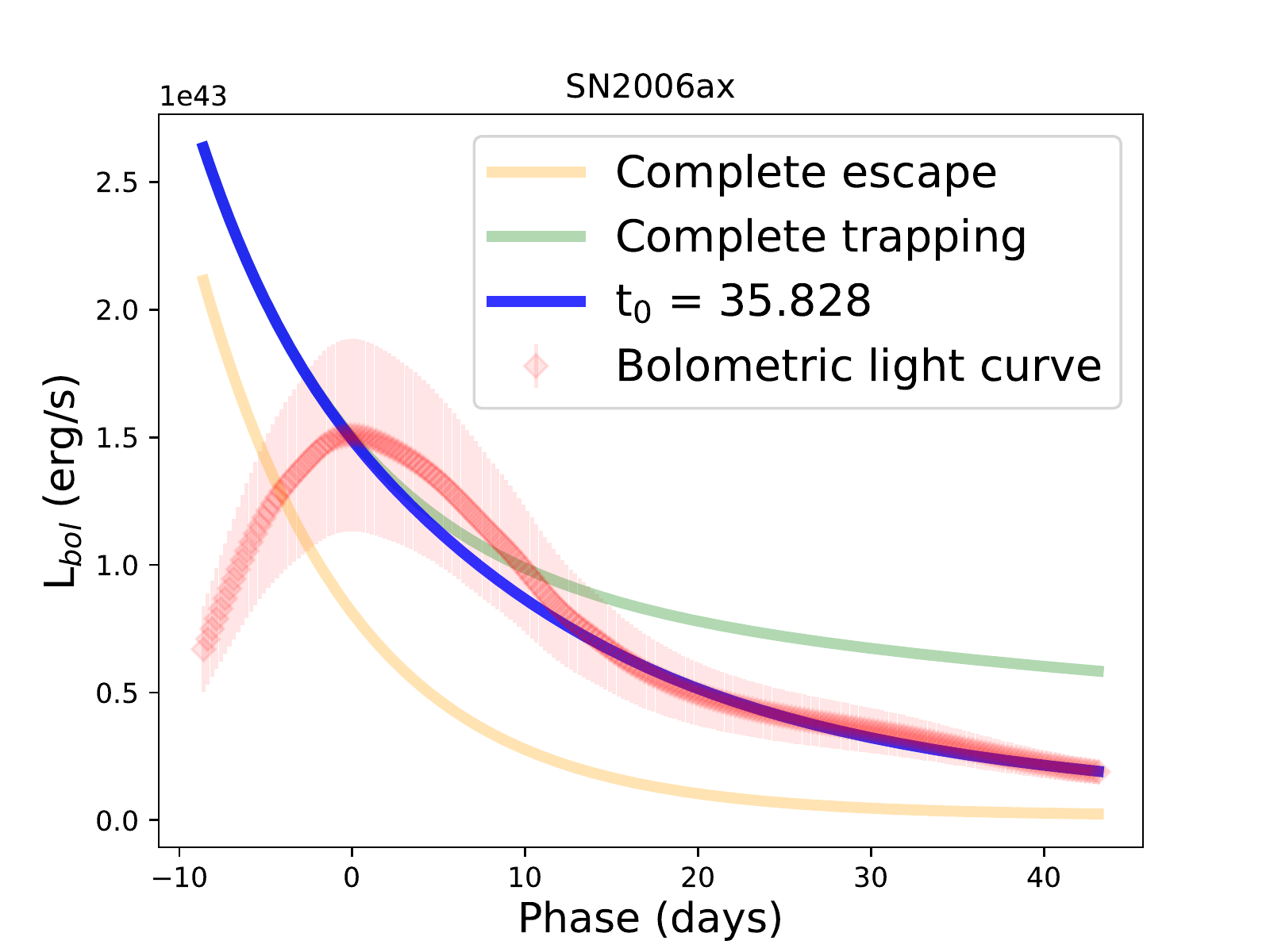}
\includegraphics[width=.2\textwidth]{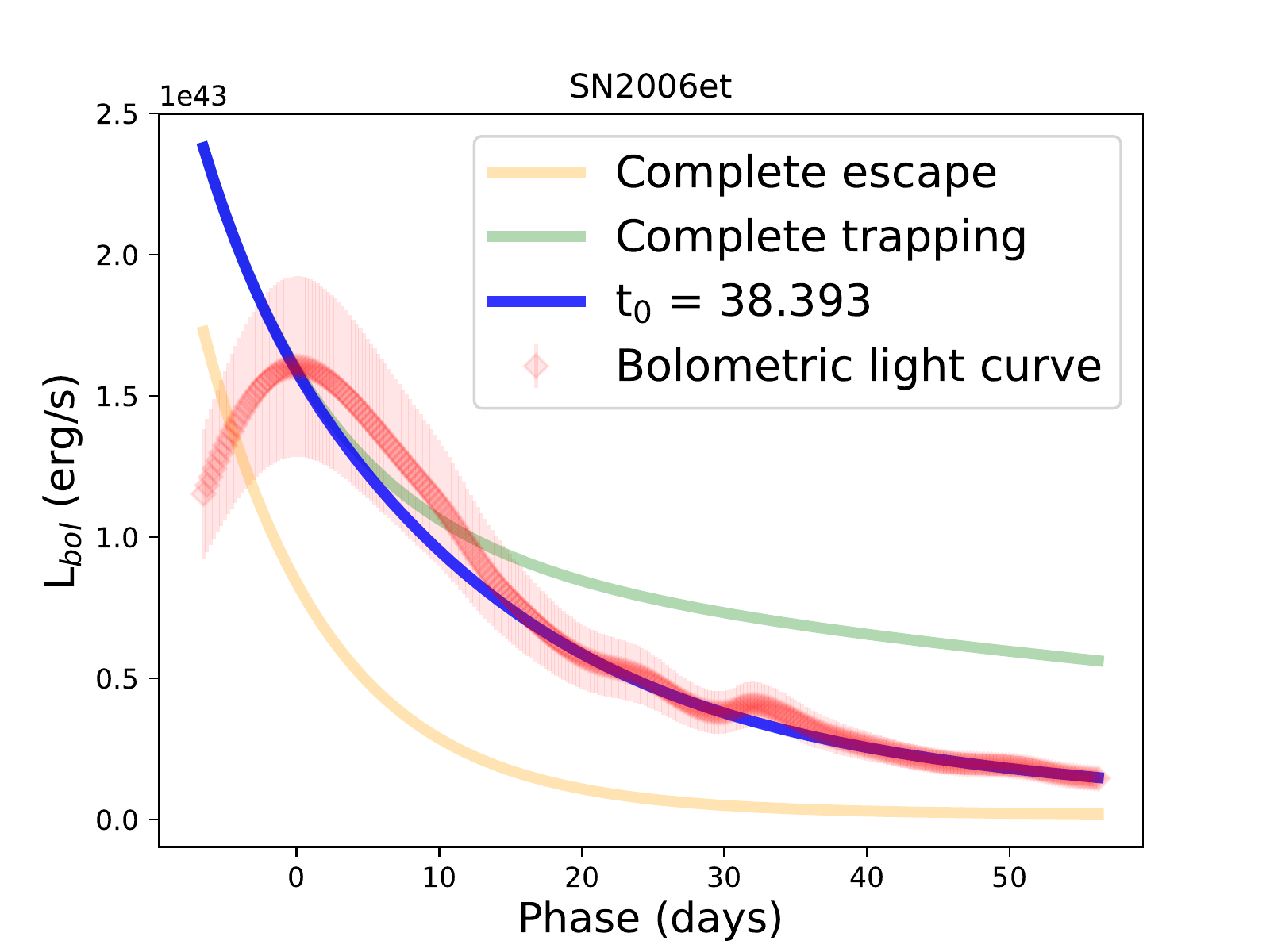}
\includegraphics[width=.2\textwidth]{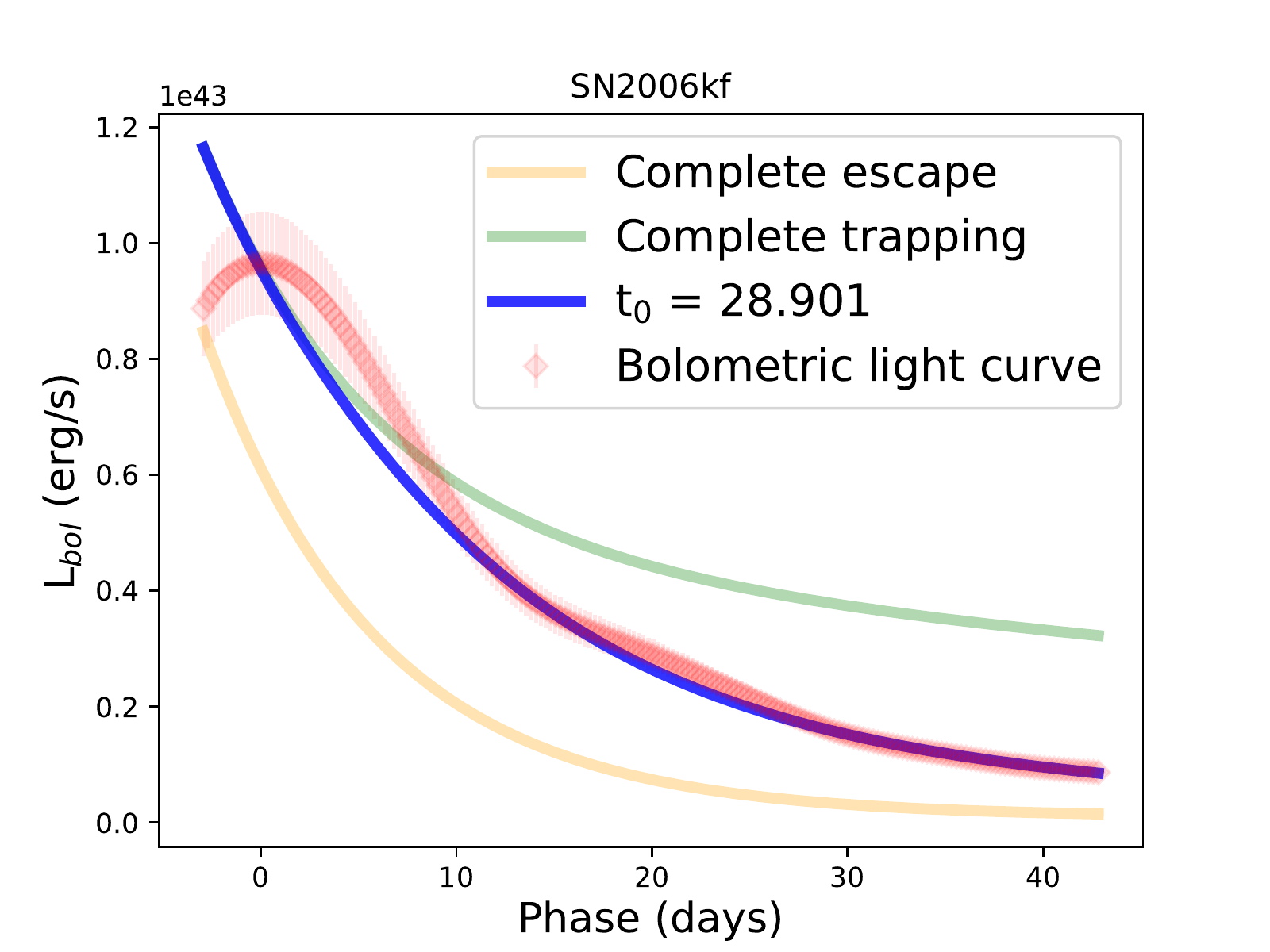}
\includegraphics[width=.2\textwidth]{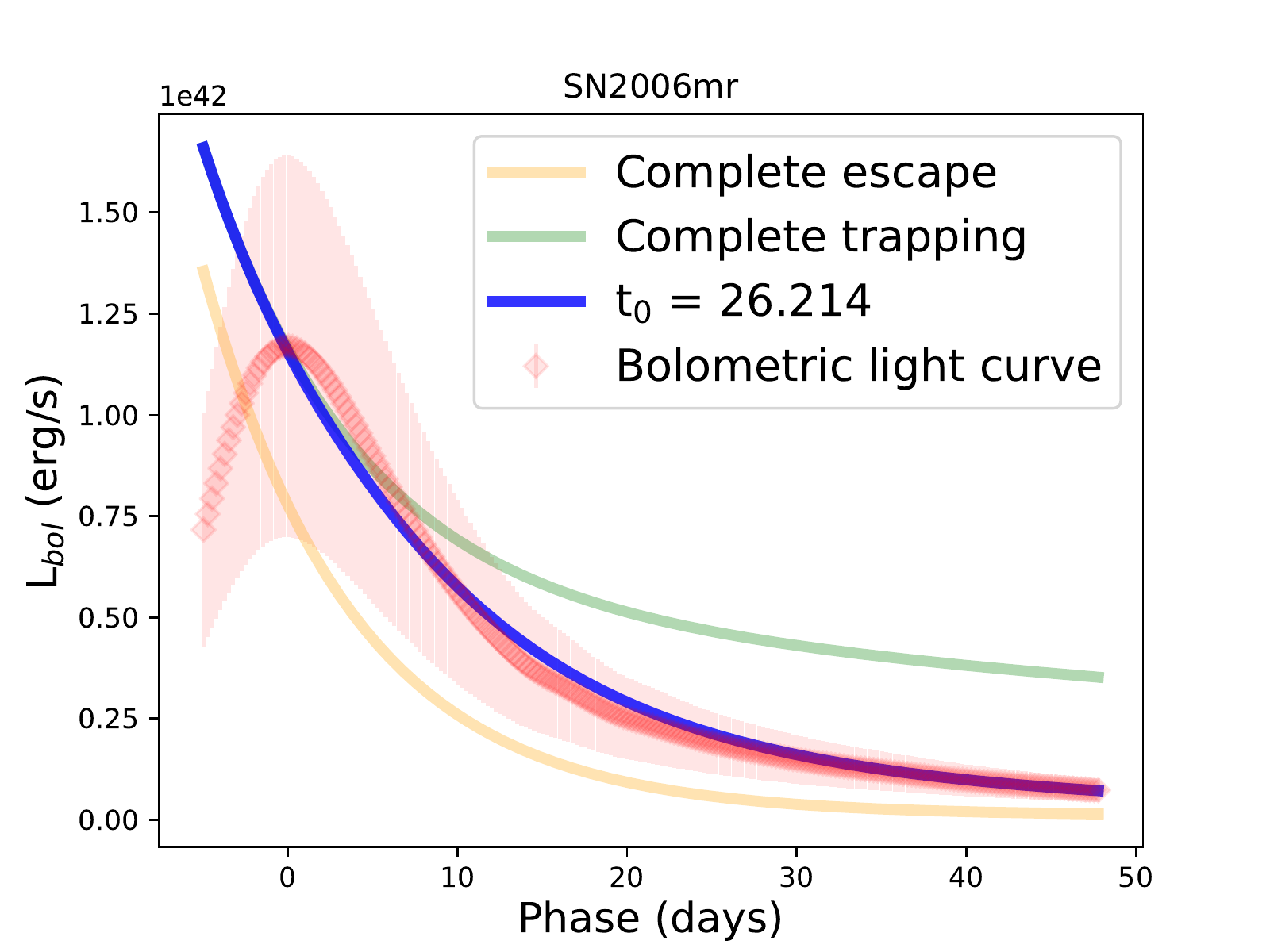}
\includegraphics[width=.2\textwidth]{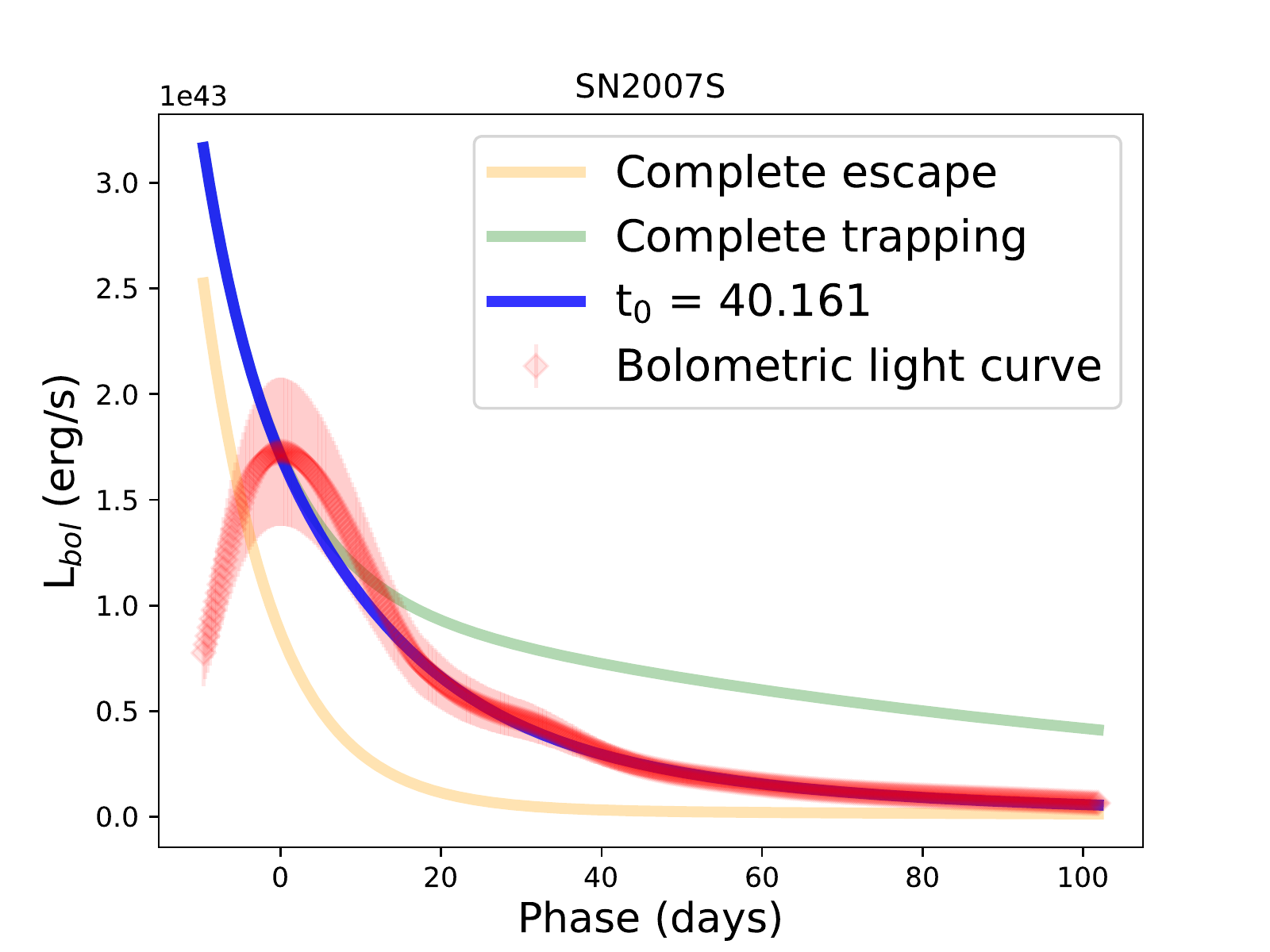}
\includegraphics[width=.2\textwidth]{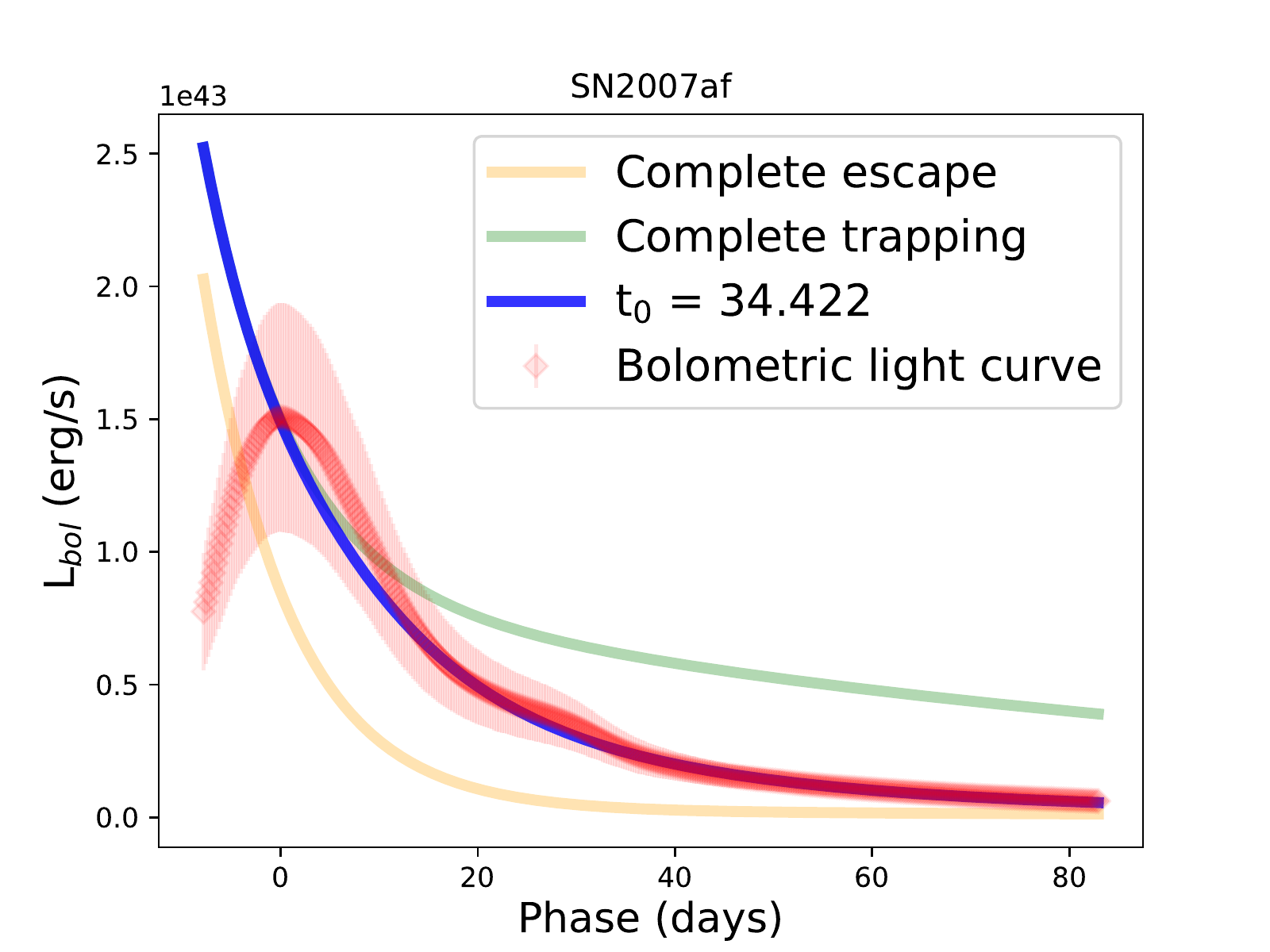}
\includegraphics[width=.2\textwidth]{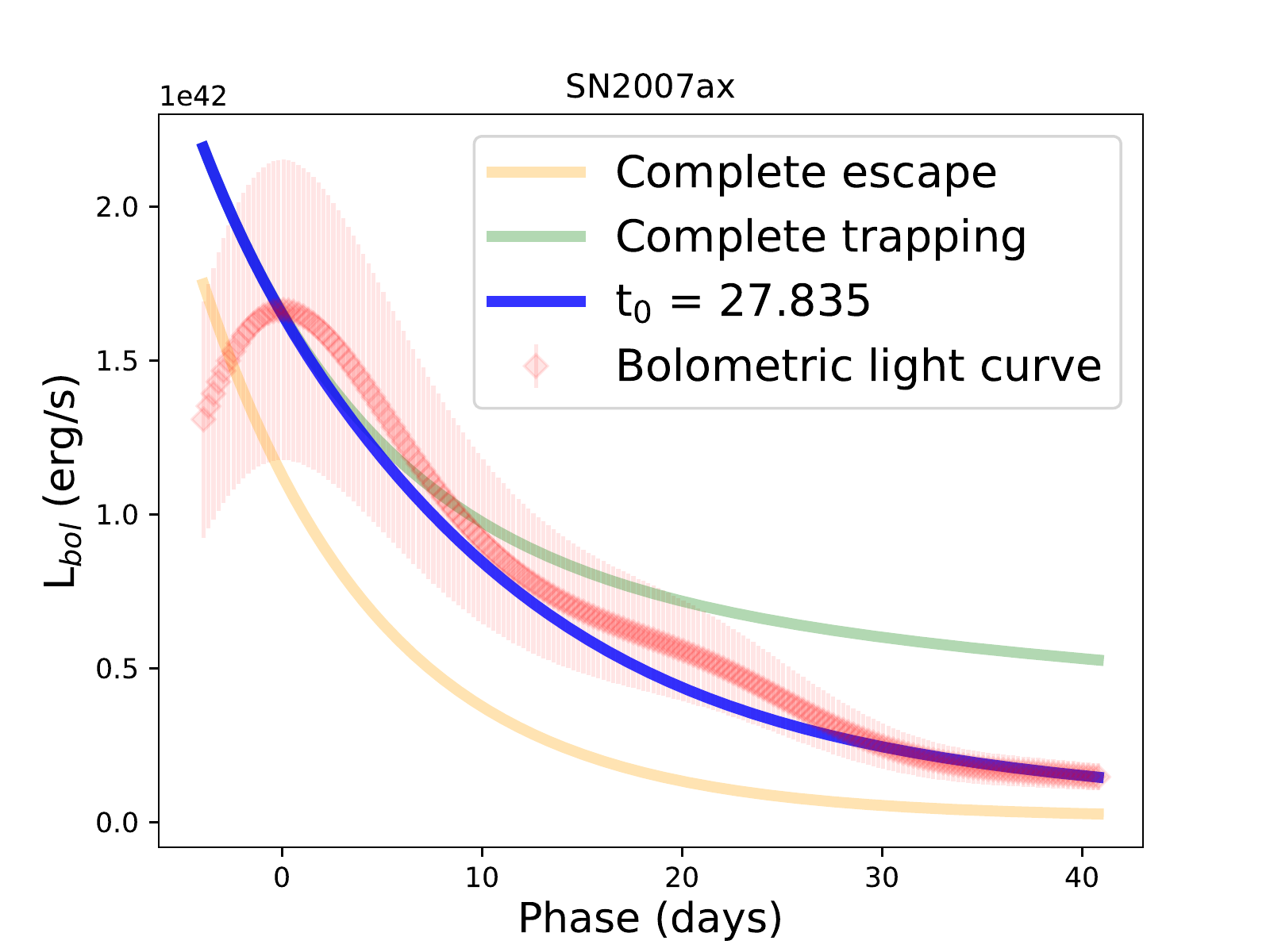}
\includegraphics[width=.2\textwidth]{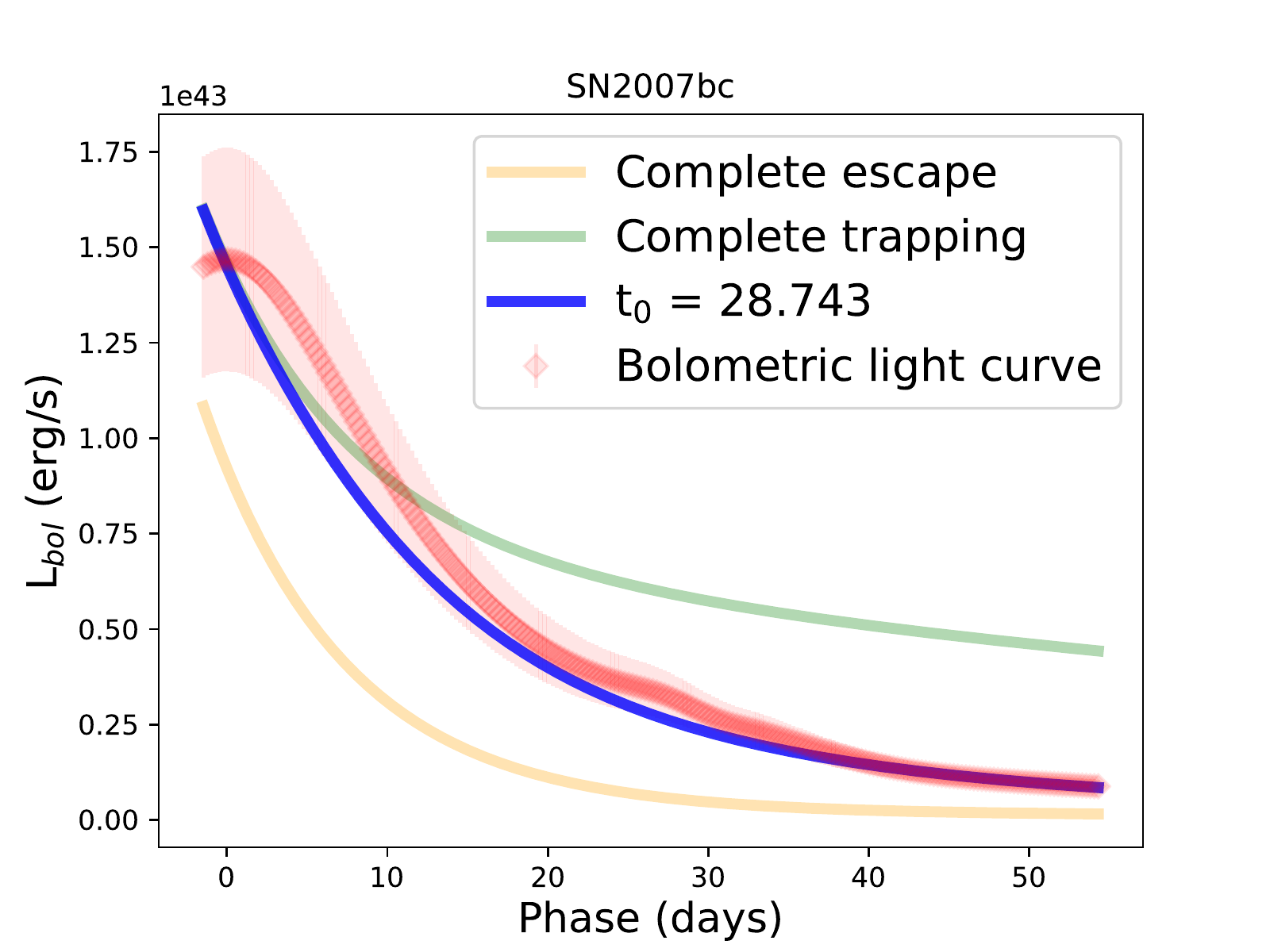}
\includegraphics[width=.2\textwidth]{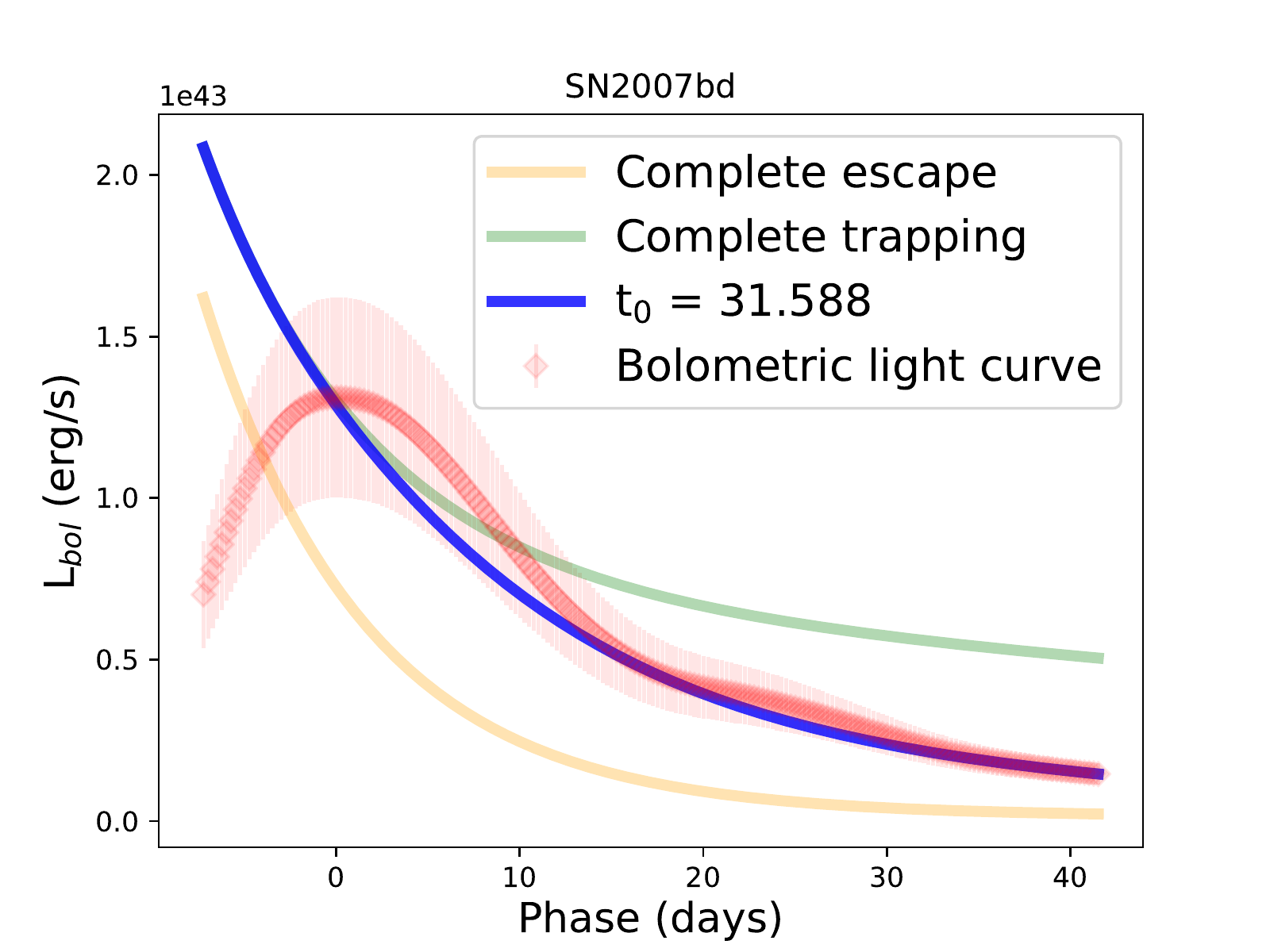}
\includegraphics[width=.2\textwidth]{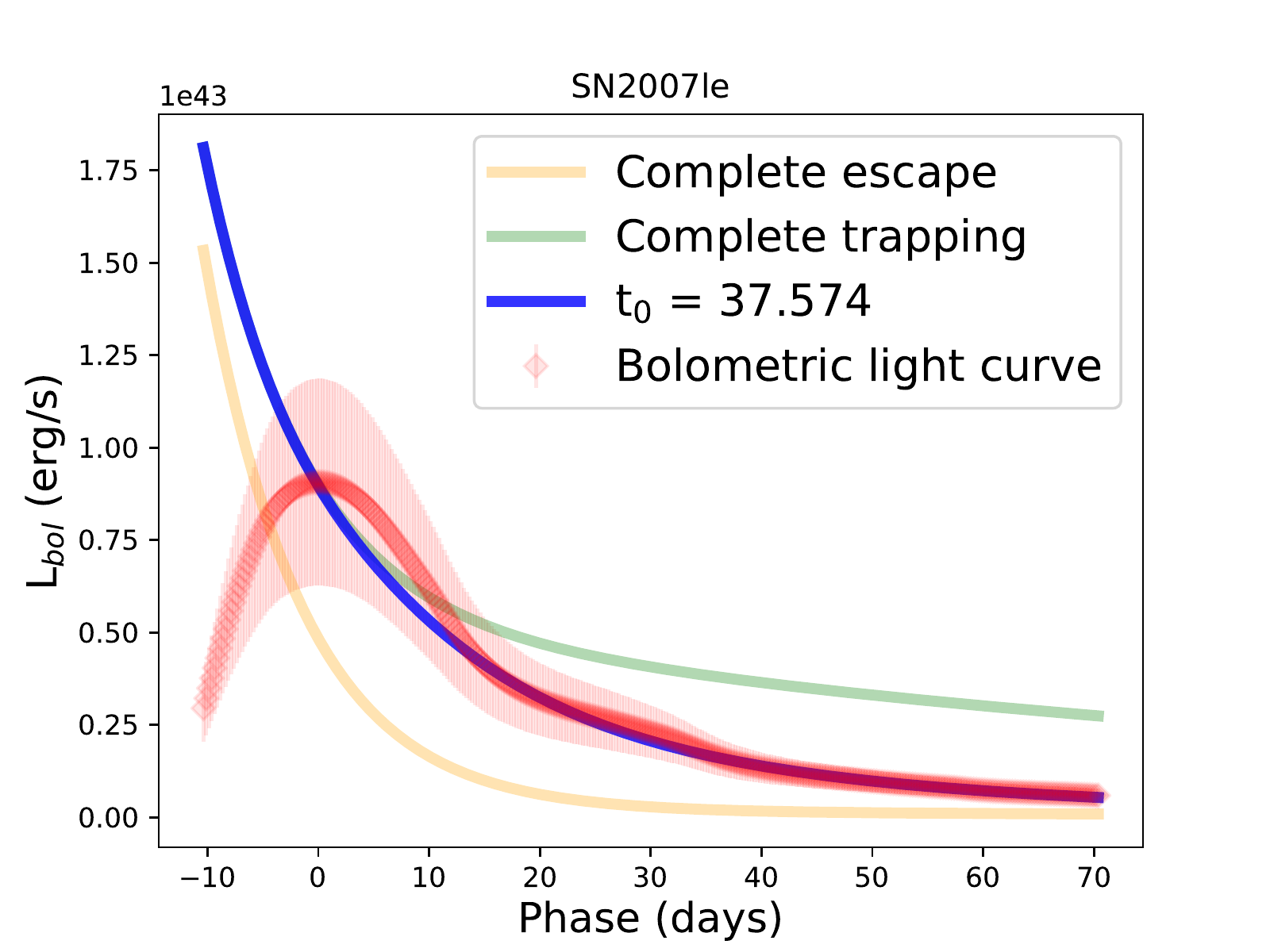}
\includegraphics[width=.2\textwidth]{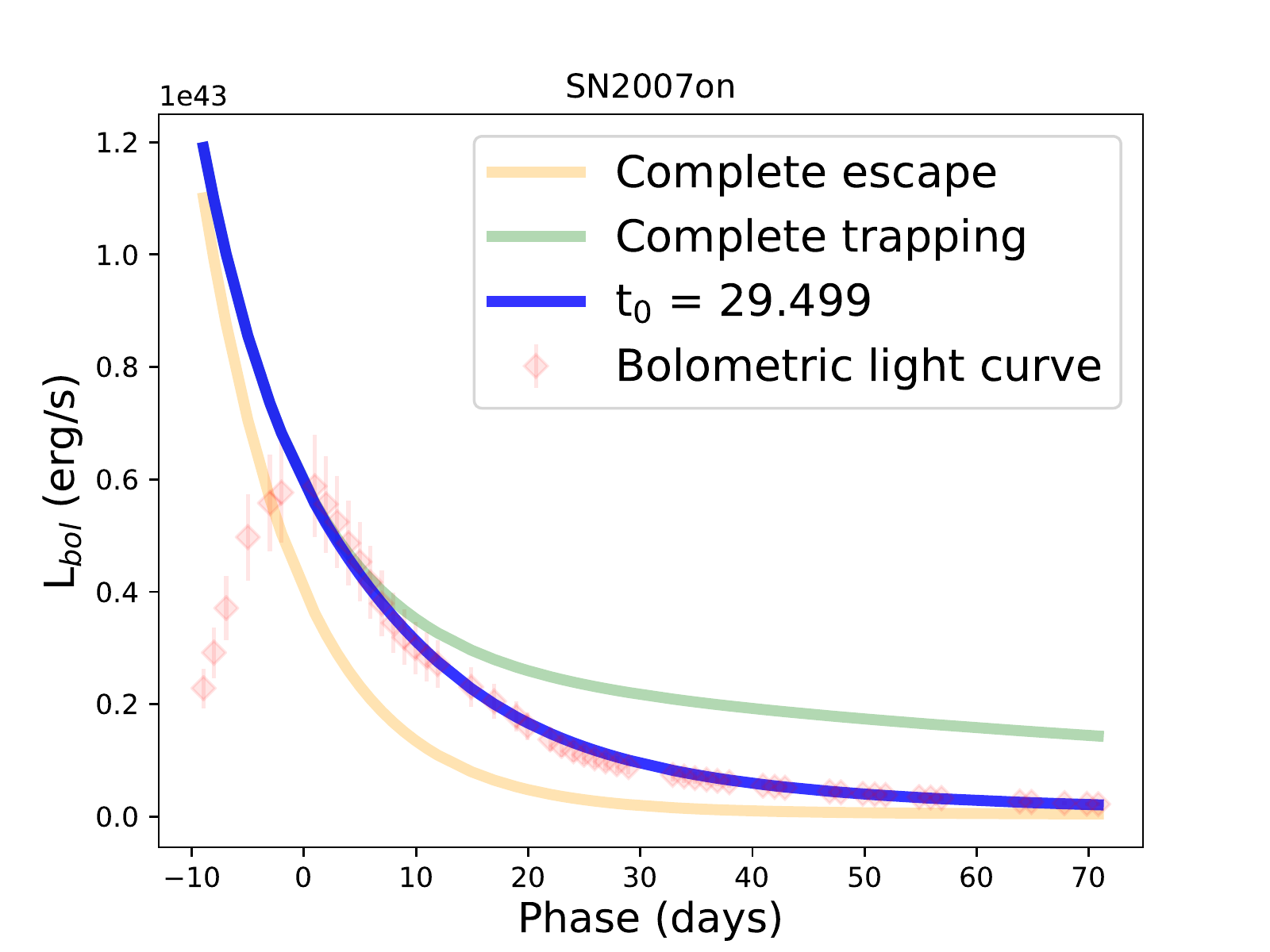}
\includegraphics[width=.2\textwidth]{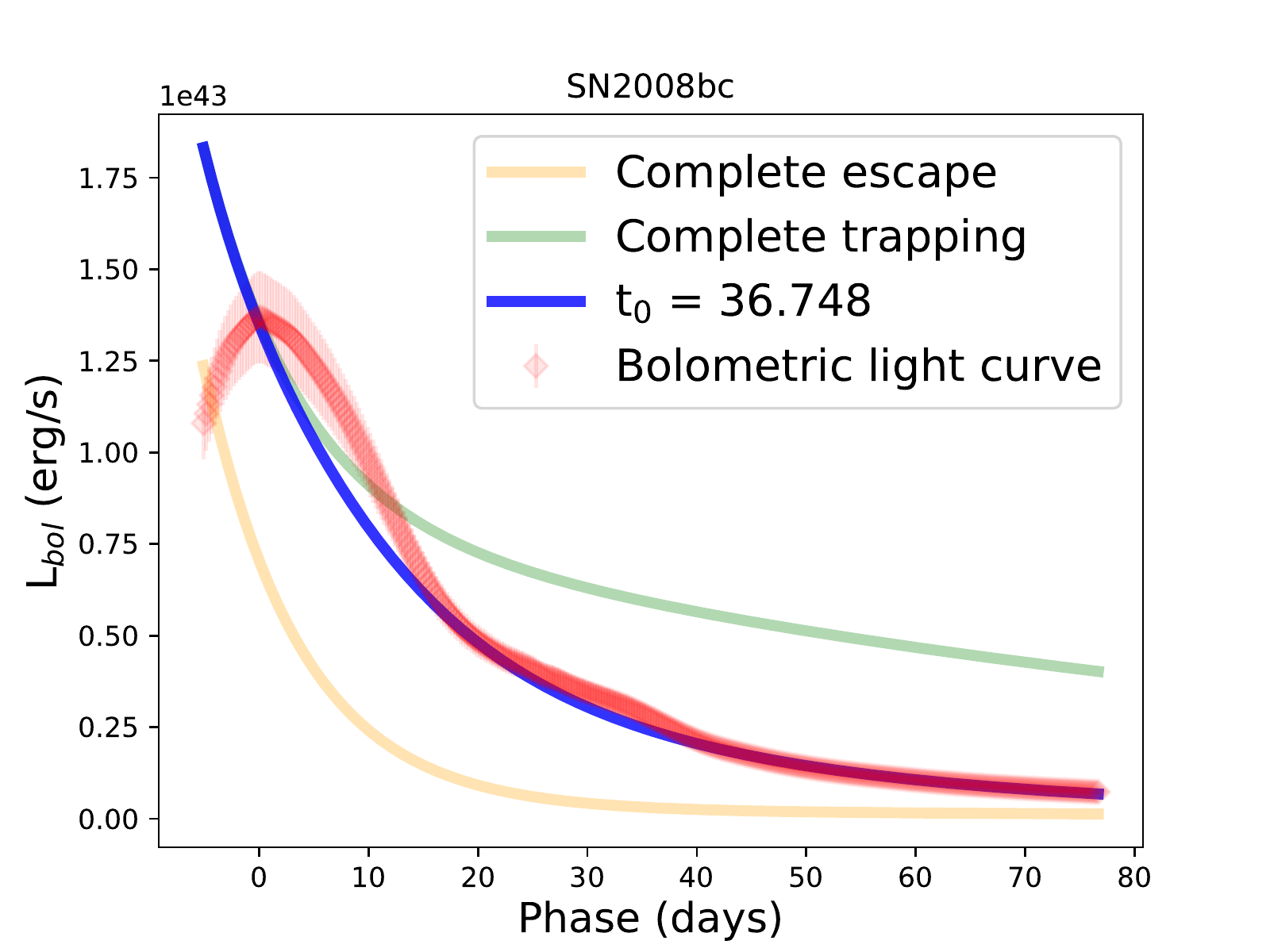}
\includegraphics[width=.2\textwidth]{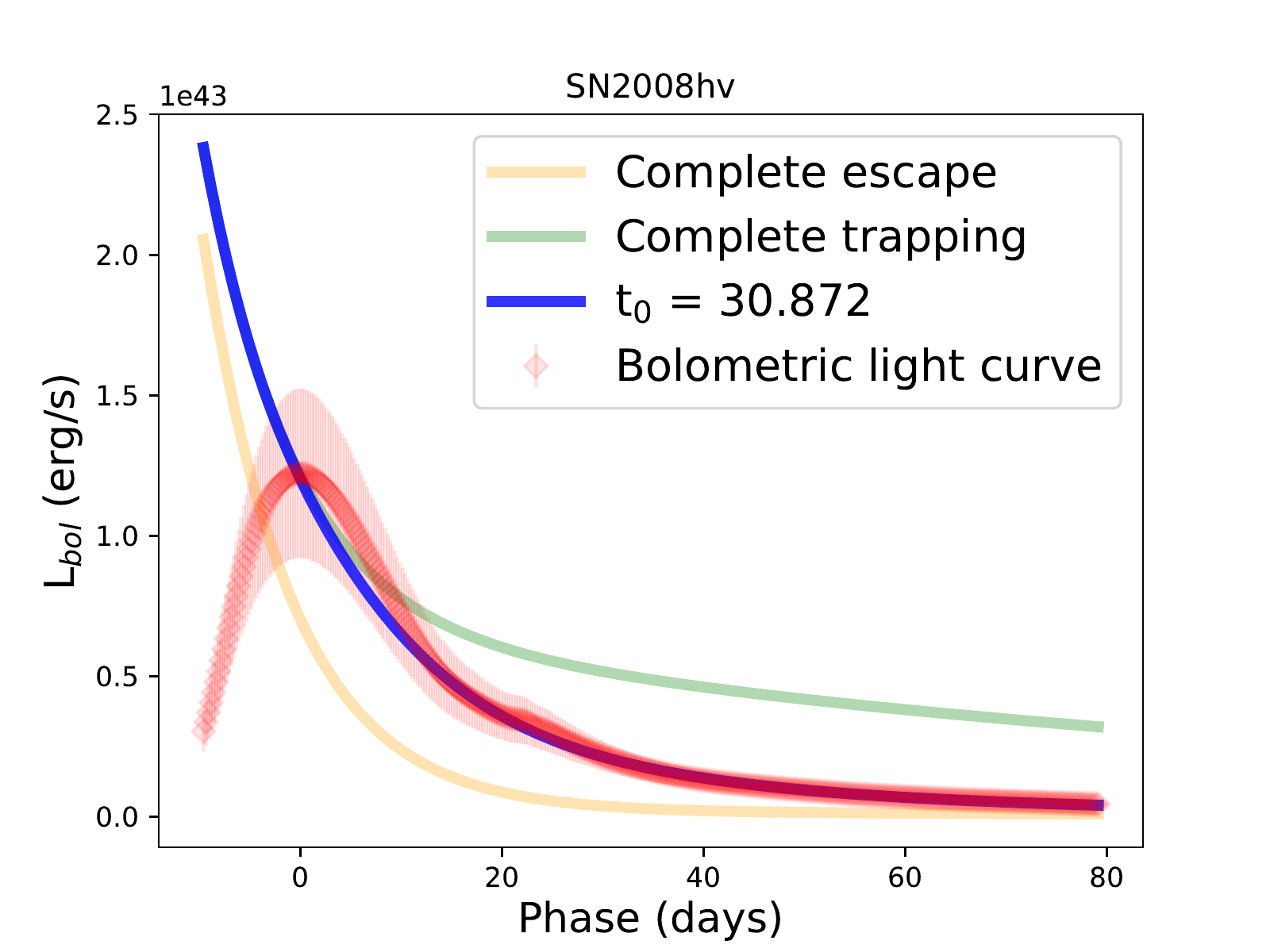}
\includegraphics[width=.2\textwidth]{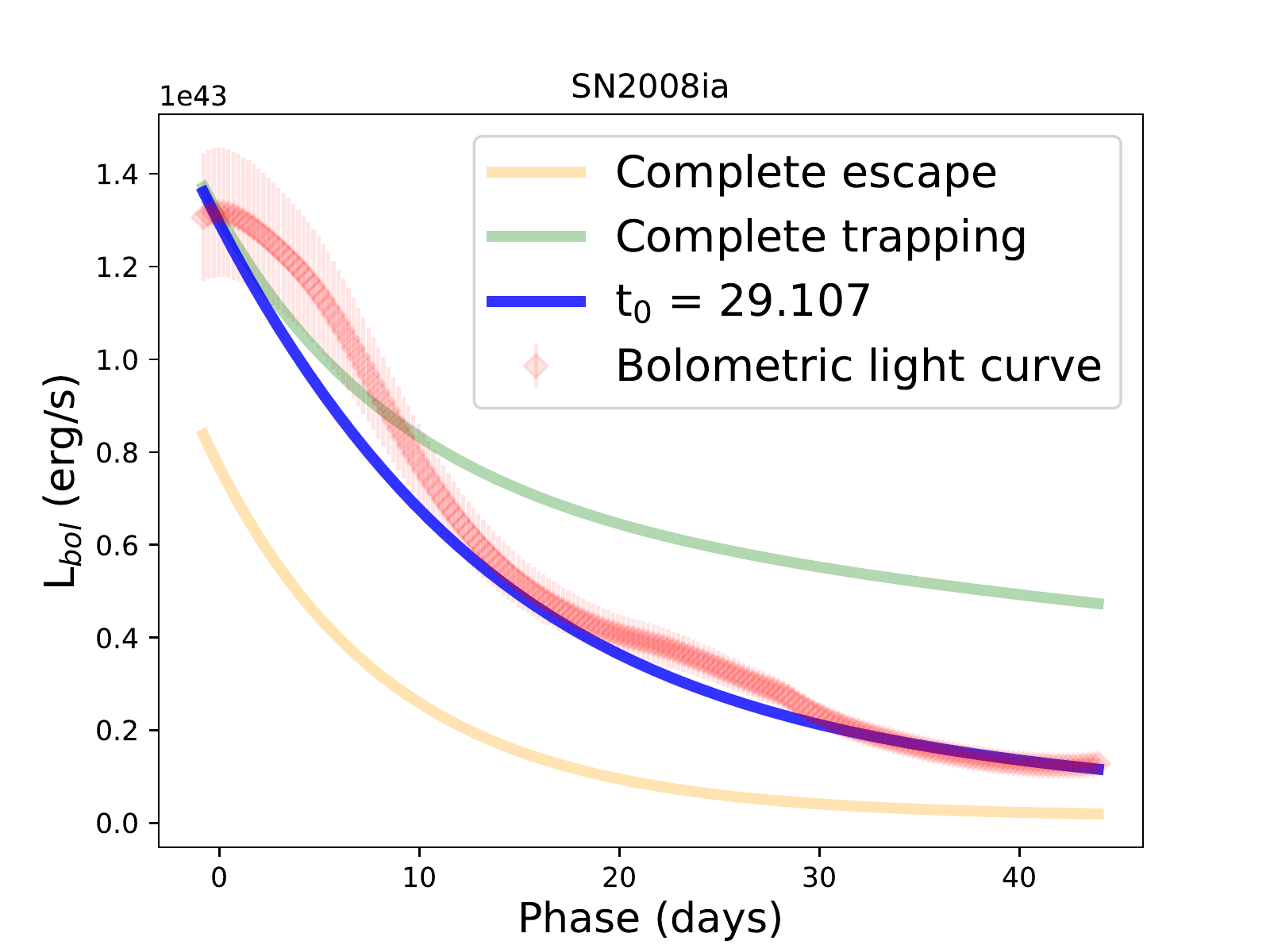}
\caption{The radioactive decay energy deposition fits to the bolometric light curve for the sample of SNe analysed in this study. For each SN the fits are presented along with the constructed light curve and the deposition function for complete trapping and complete escape of $\gamma$-rays.}
\end{figure}

\bsp	
\label{lastpage}
\end{document}

%% file: spectroscopic_log.tex
\begin{table*}
\caption{Log of spectroscopic observations.}
\begin{minipage}{70mm}
\begin{tabular}{|c|c|c|c|c|}
\hline\hline
Date & Phase (d) & Telescope + Instrument & Range (\AA) & Reference\footnote{M18:\citet{Miller2017}} \\
\hline
2016-04-05 & $-$15.3 & Gemini-North+GMOS & 3800 - 9200 & M18\\
2016-04-10 & $-$10.9 & Keck-I+LRIS & 3055 - 10411 & M18\\
2016-04-13 & $-$7.8 & LCO-2m+FLOYDS & 3300 - 9998 & M18\\
2016-04-25 & +3.7 & LCO-2m+FLOYDS & 3300 - 9999 & M18\\
2016-04-30 & +8.5 & LCO-2m+FLOYDS & 3301 - 9999 & M18\\
2016-05-21 & +29.2 & LCO-2m+FLOYDS & 4000 - 8998 & M18\\  
2016-06-03 & +41.9 & LCO-2m+FLOYDS & 4000 - 8998 & M18\\  
2016-06-11 & +49.7 & LCO-2m+FLOYDS & 4001 - 8999 & M18\\ 
2016-06-23 & +61.4 & LCO-2m+FLOYDS & 4800 - 9300 & M18 \\ 
2017-03-29 & +342.4 & Keck-I+LRIS & 3200 - 10000 & This work \\
\hline
\end{tabular}
\end{minipage}
\label{tab:spec_log}
\end{table*}

%% file: dm15_vals.tex
\begin{table}
\caption{This table contains the estimates for the decline rate in the $B$-band, $
\Delta m_{15}$, host galaxy extinction, $E(B-V)_{host}$, and the total to selective absorption, $R_V$, for the host galaxy of the SN. The $\Delta m_{15}(B)$ is calculated with a Gaussian Process (GP) fit to the observed $B$-band data and the $E(B-V)_{host}$ and $R_V$ are calculated from 
\texttt{SNooPy} fits \citep{Burns2011} to multi-band data.}
\begin{tabular}{|c|c|c|c|c|}
\hline\hline
SN & $\Delta m_{15}(B)$ & $E(B-V)_{host}$ & $\sigma$ & $R_V$\\
& (mag) & (mag)  & (mag) & \\
\hline 
SN2004eo & 1.42 & 0.128 & 0.024 & 0.9 \\
SN2004ey & 0.96 & 0.019 & 0.020 & 3.1 \\
SN2005el & 1.35 & 0.015 & 0.012 & 3.5 \\
SN2005ke & 1.78 & 0.263 & 0.033 & 0.8 \\
SN2005ki & 1.28 & 0.016 & 0.013 & 3.4 \\
SN2005M & 0.90 & 0.060 & 0.021 & 3.4 \\
SN2006ax & 1.00 & 0.016 & 0.015 & 2.9 \\
SN2006D & 1.43 & 0.134 & 0.025 & 2.5 \\
SN2006et & 0.90 & 0.254 & 0.025 & 1.9 \\
SN2006kf & 1.63 & 0.032 & 0.011 & 4.1 \\
SN2006mr & 1.83 & 0.089 & 0.039 & 2.9 \\
SN2007af & 1.17 & 0.178 & 0.024 & 2.1 \\
SN2007ax & 1.90 & 0.213 & 0.049 & 2.6 \\
SN2007bc & 1.59 & 0.207 & 0.025 & 1.8 \\
SN2007bd & 1.10 & 0.058 & 0.022 & 2.1 \\
SN2007le & 0.97 & 0.388 & 0.023 & 1.7 \\
SN2007on & 1.88 & 0.000 & 0.000 & 3.5 \\
SN2007S & 0.77 & 0.478 & 0.026 & 1.9 \\
SN2008bc & 0.85 & 0.000 & 0.000 & 3.1 \\
SN2008hv & 1.25 & 0.074 & 0.023 & 2.1 \\
SN2008ia & 1.29 & 0.066 & 0.016 & 3.8 \\
iPTF16abc & 0.91 & 0.070 & 0.016 & 3.1\\
\hline
\end{tabular}
\label{tab:ebv_rv}
\end{table}

%% file: distances.tex
\begin{table}
\caption{Distances for the SNe in our sample. These were used to calculate the absolute (pseudo)-bolometric flux.}
\begin{tabular}{|c|c|c|}
\hline\hline
SN & $\mu$ & $\sigma$\\
	& (mag) & (mag) \\
\hline
SN2004eo & 34.03 & 0.14 \\
SN2004ey & 34.01 & 0.15 \\
SN2005el & 34.05 & 0.14 \\
SN2005ke & 31.84 & 0.08 \\
SN2005ki & 34.74 & 0.11 \\
SN2005M & 35.0 & 0.09 \\
SN2006ax & 34.46 & 0.12 \\
SN2006D & 33.1 & 0.22 \\
SN2006et & 34.82 & 0.10 \\
SN2006kf & 34.78 & 0.10 \\
SN2006mr & 31.15 & 0.23 \\
SN2006X & 32.17 & 0.32 \\
SN2007af & 32.16 & 0.32 \\
SN2007ax & 32.2 & 0.27 \\
SN2007bc & 34.89 & 0.10 \\
SN2007bd & 35.73 & 0.07 \\
SN2007le & 31.88 & 0.36 \\
SN2007on & 31.45 & 0.08 \\
SN2007S & 34.07 & 0.14 \\
SN2008bc & 34.17 & 0.14 \\
SN2008fp & 32.16 & 0.32 \\
SN2008hv & 33.85 & 0.16 \\
SN2008ia & 34.97 & 0.10 \\
iPTF16abc & 35.01 & 0.09 \\
\hline 
\end{tabular}
\label{tab:dist}
\end{table}

%% file: nir_table.tex
\begin{table}
\caption{The near infrared second maximum timing for iPTF16abc.}
\begin{tabular}{|c|c|c|c|}
\hline\hline
Filter & $t_2$ \\
 &  (d)\\
 \hline
Y & 34.4 $\pm$ 1.23\\
J & 34.2 $\pm$ 1.23 \\
H &  33.2 $\pm$ 1.4 \\
\hline
\end{tabular}
\label{tab:nir}
\end{table}

%% file: nickel_mass.tex
\begin{table*}
\centering
\caption{The derived \Mni\, for the sample of SNe with a measured $t_0$ under two different assumptions on the rise time of the SNe. From the values below we find that the inferred \Mni\, is within the errors even with individual rise times for each SN.}
\begin{tabular}{|c|c|c|c|c|c|c|}
\hline\hline
SN & L$_{max}$ & $\sigma$ & M$_{\mathrm{^{56}Ni}}$ (fixed rise) & $\sigma$ &  M$_{\mathrm{^{56}Ni}}$ (variable rise) & $\sigma$\\
& 10$^{43}$ erg/s & 10$^{43}$ erg/s & M$_{\odot}$ & M$_{\odot}$ & M$_{\odot}$ & M$_{\odot}$\\
\hline 
SN2004eo & 1.08 & 0.14 & 0.54 & 0.11 & 0.44 & 0.08 \\
SN2004ey & 1.27 & 0.21 & 0.63 & 0.14 & 0.59 & 0.12 \\
SN2005el & 1.20 & 0.18 & 0.6 & 0.13 & 0.50 & 0.10 \\
SN2005ke & 0.31 & 0.04 & 0.15 & 0.03 & 0.11 & 0.02 \\
SN2005ki & 1.19 & 0.37 & 0.6 & 0.21 & 0.51 & 0.17 \\
SN2005M & 1.33 & 0.36 & 0.67 & 0.21 & 0.63 & 0.18 \\
SN2006ax & 1.51 & 0.38 & 0.75 & 0.22 & 0.69 & 0.19 \\
SN2006D & 1.35 & 0.32 & 0.68 & 0.19 & 0.55 & 0.15 \\
SN2006et & 1.60 & 0.32 & 0.80 & 0.20 & 0.76 & 0.17 \\
SN2006kf & 0.97 & 0.09 & 0.48 & 0.08 & 0.37 & 0.06 \\
SN2006mr & 0.12 & 0.05 & 0.06 & 0.03 & 0.04 & 0.02 \\
SN2007af & 1.51 & 0.43 & 0.75 & 0.24 & 0.66 & 0.20 \\
SN2007ax & 0.17 & 0.05 & 0.08 & 0.03 & 0.06 & 0.02 \\
SN2007bc & 1.47 & 0.29 & 0.73 & 0.18 & 0.57 & 0.13 \\
SN2007bd & 1.31 & 0.31 & 0.66 & 0.18 & 0.59 & 0.15 \\
SN2007le & 0.91 & 0.28 & 0.45 & 0.16 & 0.42 & 0.14 \\
SN2007on & 0.58 & 0.03 & 0.29 & 0.05 & 0.21 & 0.03 \\
SN2007S & 1.73 & 0.35 & 0.86 & 0.22 & 0.84 & 0.19 \\
SN2008bc & 1.37 & 0.12 & 0.68 & 0.12 & 0.65 & 0.09 \\
SN2008hv & 1.22 & 0.30 & 0.61 & 0.18 & 0.52 & 0.14 \\
SN2008ia & 1.32 & 0.14 & 0.66 & 0.12 & 0.56 & 0.09 \\
iPTF16abc & 1.22 &0.20 & 0.61 & 0.14 & 0.55 & 0.11 \\
\hline
\end{tabular}
\label{tab:ni}
\end{table*}

%% file: tab_t0_16abc_withrise.tex
\begin{table*}
\caption{The values of the  transparency timescales for a sample of SNe from the literature with sufficient data. The first estimate assume a rise time of 19\,d whereas the second assumes a rise time that varies as a function of the measured $\Delta m_{15}(B)$. An additional systematic uncertainty of $\sim$ 2 days should be added in quadrature to the error on $t_0$ due to the uncertainty in knowing the rise time. The values from the two different estimates are comparable within the systematic errors.}
\centering
\begin{tabular}{|c|c|c|c|c|c|c|}
\hline\hline
SN  & $t_0$ (fix rise) & err & $t_0$ (variable rise) & err\\
& (days) & (days) & (days) & (days) \\
\hline
SN2004eo & 35.34 & 0.06 & 36.67 & 0.08 \\
SN2004ey & 34.64 & 0.03 & 34.90 & 0.03 \\
SN2005el & 28.38 & 0.03 & 29.49 & 0.05 \\
SN2005ke & 28.87 & 0.02 & 31.35 & 0.05 \\
SN2005ki & 29.85 & 0.04 & 30.69 & 0.06 \\
SN2005M & 38.01 & 0.14 & 38.22 & 0.14 \\
SN2006ax & 35.51 & 0.08 & 35.83 & 0.07 \\
SN2006D & 28.47 & 0.02 & 29.89 & 0.05 \\
SN2006et & 38.18 & 0.09 & 38.39 & 0.09 \\
SN2006kf & 27.62 & 0.04 & 28.90 & 0.06 \\
SN2006mr & 24.54 & 0.01 & 26.21 & 0.02 \\
SN2007af & 33.50 & 0.06 & 34.42 & 0.08 \\
SN2007ax & 26.12 & 0.02 & 27.84 & 0.04 \\
SN2007bc & 27.28 & 0.05 & 28.74 & 0.05 \\
SN2007bd & 31.21 & 0.01 & 31.59 & 0.01 \\
SN2007le & 37.12 & 0.07 & 37.57 & 0.08 \\
SN2007on & 26.63 & 0.01 & 29.08 & 0.06 \\
SN2007S & 40.18 & 0.09 & 40.16 & 0.09 \\
SN2008bc & 36.60 & 0.05 & 36.75 & 0.05 \\
SN2008hv & 29.88 & 0.04 & 30.87 & 0.07 \\
SN2008ia & 28.42 & 0.22 & 29.11 & 0.24 \\
iPTF16abc & 39.50 & 0.21 & 39.76 & 0.22\\
\hline
\end{tabular}
\label{tab:t0}
\end{table*}

%% file: photometry_de.tex
\begin{table*}
\caption{Nebular Phase photometry with DECam}
\begin{tabular}{|c|c|c|c|c|}
\hline\hline
MJD & Phase (d) & filter &  magnitude & $\sigma$ \\
& & & (mag) & (mag)\\
\hline
57818.30 & 320.20 & desg & 22.249 & 0.053  \\
57818.30 & 320.20& desr &  23.323 & 0.053 \\
57814.39 & 316.29 & desz & 22.501 & 0.191 \\
\hline
\end{tabular}
\label{tab:phot_decals}
\end{table*}

%% file: thalf_vals.tex
\begin{table*}
\caption{The values of the (pseudo-) bolometric peak luminosities and the transparency timescales for a sample of SNe from the literature with sufficient data to measure both quantities. An additional systematic uncertainty of $\sim$ 2 days should be added in quadrature to the error on $t_0$ due to the uncertainty in knowing the rise time.}
\centering
\begin{tabular}{|c|c|c|c|c|c|c|}
\hline\hline
SN & t$_{+1/2}$ & $t_0$ (variable rise) & err\\
& (d)  & (d) & (d) \\
\hline
SN2004ey & 13.29 & 34.9 & 0.03 \\
SN2005el & 11.09 & 29.49 & 0.05 \\
SN2005ke & 11.14 & 31.35 & 0.05 \\
SN2005ki & 12.24 & 30.69 & 0.06 \\
SN2005M & 14.73 & 38.22 & 0.14 \\
SN2006D & 11.13 & 29.89 & 0.05 \\
SN2006et & 14.59 & 38.39 & 0.09 \\
SN2006kf & 11.00 & 28.9 & 0.06 \\
SN2006mr & 9.62 & 26.21 & 0.02 \\
SN2007af & 13.09 & 34.42 & 0.08 \\
SN2007ax & 11.39 & 27.84 & 0.04 \\
SN2007bc & 12.85 & 28.74 & 0.05 \\
SN2007bd & 12.60 & 31.59 & 0.01 \\
SN2007on & 9.11 & 29.08 & 0.06 \\
SN2007S & 14.97 & 40.16 & 0.09 \\
SN2008bc & 14.58 & 36.75 & 0.05 \\
SN2008hv & 11.92 & 30.87 & 0.07 \\
SN2008ia & 11.84 & 29.11 & 0.24 \\
iPTF16abc & 14.14 & 39.76 & 0.09 \\
\hline
\end{tabular}
\label{tab:thalf}
\end{table*}